\documentclass{article}
\input boxedeps.tex
\SetRokickiEPSFSpecial  
\HideDisplacementBoxes

\newcommand{\D}{\overline{D}}
\newcommand{\da}{\dagger}  
\newcommand{\be}{\begin{equation}}
\newcommand{\eq}{\end{equation}}
   
\newcommand{\Tr}{{\rm \, Tr \!}}    
\newcommand{\newl}{l}               
\newcommand{\newtau}{t}             
\newcommand{\dm}{{\cal M}}          

\begin{document}

%

\mbox{}\hfill DAMTP-1999-107\\
\vspace{10mm}
\begin{center}
{\Huge  Glueballs on a transverse lattice}
\vspace{30mm}

{\Large S. Dalley${}^{*}$ and  B. van de Sande${}^{**}$}\\
\vspace{10mm}

{\em
${}^*$Department of Applied Mathematics and Theoretical Physics, \\
 Silver Street, Cambridge CB3 9EW, England
\vspace{5mm}

${}^{**}$Geneva College,\\
3200 College Ave., Beaver Falls, PA 15010
}
\vspace{30mm}

\end{center}
\begin{abstract}

Accurate non-perturbative calculations of glueballs 
are performed using light-front quantised $SU(N)$ gauge theory,
to leading order of the $1/N$ expansion. Based on early work of
Bardeen and Pearson, disordered gauge-covariant link variables $M$ on
a coarse transverse lattice are used to approximate the physical
gauge
degrees of freedom.  Simple energetics imply that, at lattice spacings
of order the inverse QCD scale, the effective light-front Hamiltonian
can be expanded in gauge-invariant powers of $M$: a colour-dielectric
expansion.  This leads to a self-consistent constituent structure of
boundstates.  We fix the couplings of this expansion by optimising
Lorentz covariance of low-energy eigenfunctions.  To lowest
non-trivial order of the expansion, we have found a one-parameter
trajectory of couplings that enhances Lorentz covariance.  On this
trajectory the masses of nearly-covariant glueball states exhibit
approximate scaling, having values consistent with large-$N$
extrapolations of continuum results from other methods. There is very
little variation with $N$ in pure Yang-Mills theory:  the lightest
glueball mass changes by only a few percent between $SU(3)$ and
$SU(\infty)$.  The corresponding light-front wavefunctions show an
unconventional structure.
We also examine restoration of rotational invariance in
the heavy-source potential.

\end{abstract}

\newpage
\baselineskip .2in


\section{Introduction}
\label{intro}

There are few, if any, efficient methods for tackling relativistic
strongly-bound states in generic
four-dimensional gauge theories.
The canonical example is QCD, where non-perturbative theoretical calculations
have rarely been ahead of experiment.
Future progress in particle physics is likely
to hinge upon a detailed theoretical understanding of these questions. 
This has led some theorists to develop 
Hamiltonian quantisation on a light-front \cite{dirac,review}.
In the presence of
suitable high-energy cut-offs, the light-front vacuum state is trivial and  
wavefunctions built upon it are Lorentz boost-invariant.

In particular, Brodsky and Lepage \cite{stan} and Pauli and Brodsky
\cite{dlcq} have urged the development of the light-front quantisation
of QCD (LFQCD). More recently, Wilson {\em et al.}\ \cite{wilson1} have
clarified the physical principles underlying LFQCD and suggested a
weak-coupling calculational framework. In this paper we develop an
alternative framework which appears promising: a light-front
quantisation of lattice gauge theory \cite{bard1}.  Calculations that
we have already performed with this method \cite{dv0,dv1,dv2}, for
non-Abelian gauge theories in $2+1$ dimensions, were surprisingly
successful in comparison to results from traditional Euclidean lattice
path integral simulations (ELMC) \cite{teper1}.  They have also
produced new results in the form of the light-front wavefunctions, the
starting point for investigation of virtually any physically
interesting observable.  Encouraged by this success, we 
investigate here the 
glueballs and heavy-source potential in $3+1$-dimensional 
gauge theory without fermions.

The well-known triviality of the light-front
vacuum may be reconciled with the 
conventional picture of a complicated QCD vacuum.  In
light-front co-ordinates, vacuum structure is carried by an isolated
set of (infinitely) high energy modes, which are removed by the
cut-off.  According to standard lore, one would expect that
appropriate renormalisation of the Hamiltonian and other observables
would recover the information excised by the high-energy cut-off.  In
this way, effects normally associated with the vacuum, such as
spontaneous symmetry breaking, must appear explicitly in the
Hamiltonian.  But the non-perturbative renormalisation group (RG)
formalism to systematically compute the necessary counter-terms does
not yet exist. Efforts are being made to formulate a perturbative
light-front RG for QCD \cite{ohio}, but the full vacuum structure must
necessarily appear via counter-terms that are non-analytic in the perturbative
couplings. Faced with the problem of finding non-perturbatively
renormalised
Hamiltonians, we will use symmetry as our guide.

The first step is to choose the most general set of Hamiltonians which
respect symmetries of the theory that are not violated by the
cut-offs.  
After truncating this set according to some reasonable criteria,
we non-perturbatively test the low-energy eigenfunctions and
eigenvalues for restoration of the symmetries violated by the
cut-offs.  These tests, rather than RG transformations, are used to
explore the space of Hamiltonians.  Since we maintain gauge
invariance,
the symmetries in question involve only Lorentz covariance.  
We will remove all but one cut-off and find that  
our truncated space of Hamiltonians contains
a unique one-parameter trajectory ${\cal T}_s$
on which Lorentz covariance is greatly enhanced.
The validity of ${\cal T}_s$ is confirmed by the approximate 
scaling of low-energy physical
quantities along this trajectory.  ${\cal T}_s$
will be used as the basis for extracting cut-off independent results.
This is a first-principles approach, since no data are 
taken from experiment (aside from the overall scale in QCD).

Apart from Lorentz invariance, we must address the issue of gauge
invariance. One knows that gauge invariance can be maintained with 
lattice cut-offs and compact variables \cite{wilson2,suss}. 
However, in light-front coordinates a high-energy lattice cut-off can only be 
applied to the transverse directions;  the null direction on the
initial surface must remain continuous.
Also, if one were to employ the usual
formulation of lattice gauge theory with degrees of freedom in
$SU(N)$, it is not straightforward to identify the independent degrees
of freedom that are essential for canonical Hamiltonian quantisation.
The tricks of `equal-time' Hamiltonian lattice gauge theory in
temporal gauge \cite{suss} do not carry over to light-front
Hamiltonian lattice gauge theory in any convenient way \cite{griffin}.
One does not have to choose lattice variables in $SU(N)$ however.  It
was noted long ago by Bardeen and Pearson \cite{bard1}, 
that lattice
variables $M$ in the space of all complex $N \times N$ matrices were
physically more appropriate on a coarse lattice. Gauge invariance is
maintained and it is straightforward to identify the independent
degrees of freedom. The penalty is that one is too far from the
continuum to use weak-coupling perturbation theory. But in the case of
light-front Hamiltonians, this may be of limited use anyway.  Bardeen {\em
et al.}\ \cite{bard2} suggested that on coarse lattices one could
expand the light-front Hamiltonian in powers of $M$ (a kind of
strong-coupling expansion).  The validity of this expansion is subtle
however, since it rests on the dynamical properties of
light-front Fock states in this kind of theory, in particular the
weakness of couplings between sectors with different number of partons
\cite{igor}.

Expanding the light-front Hamiltonian of $SU(N)$ gauge theory to
leading non-trivial order in $M$ and $1/N$, we find the trajectory
of couplings ${\cal T}_s$ mentioned above. 
In the region of coupling space we are able to investigate, 
the transverse lattice spacing on ${\cal T}_s$
is found to be always greater than about
0.65\,fm.
However, Lorentz covariance and scaling are present to sufficient
accuracy that we
can make direct estimates of the continuum values of low-energy
observables.
\footnote{As with
other `strong-coupling' approaches, the question of the formal
connection to the
continuum limit is not directly
answered.} 

The organisation of the paper is as follows.
In the first part of the present work, we give a 
systematic treatment of tranvserse lattice gauge theory.
In Section~\ref{construct} we construct generic
light-front Hamiltonians on the transverse lattice and introduce
approximation schemes for studying them; in particular, we review the
`colour-dielectric' expansion in powers of $M$ that leads to a
constituent picture of boundstates.  By working to leading order in
$1/N$, the transverse dynamics dimensionally reduce (for a coarse
lattice) \cite{dv2}, the problem becoming mathematically equivalent to a
$1+1$-dimensional gauge theory with adjoint matter \cite{igor}.  As
well as the lattice cut-off, which can only be used in transverse
spatial directions, we also employ null-plane boundary conditions,
specifically Discrete Light-Cone Quantisation (DLCQ) \cite{dlcq}, and
Tamm-Dancoff cut-offs for transverse degrees of freedom.  The latter
gauge-invariant cut-offs can be removed by  extrapolation to give
finite answers for fixed transverse lattice spacing, 
the only remaining cut-off.

The renormalisation of the theory at fixed lattice spacing is then 
accomplished by
optimising Lorentz covariance, discussed in Section~\ref{couplings}. 
This amounts to a choice of `metric' in the space of Hamiltonians,
based on explicit calculations of observables.
Results from our calculations are
presented in Section~\ref{results}; we will not describe in any
detail the calculational techniques, but refer the reader 
to our previous papers.
Some preliminary results from the glueball  calculations 
were presented in Ref.~\cite{dv3}, but the new material here
is more accurate and extensive, including a thorough
analysis of the space of couplings, the glueball masses and
wavefunctions, and the heavy-source system.

\begin{figure}
\centering
$\displaystyle\frac{M}{\sqrt{\sigma}}$\hspace{5pt}
\BoxedEPSF{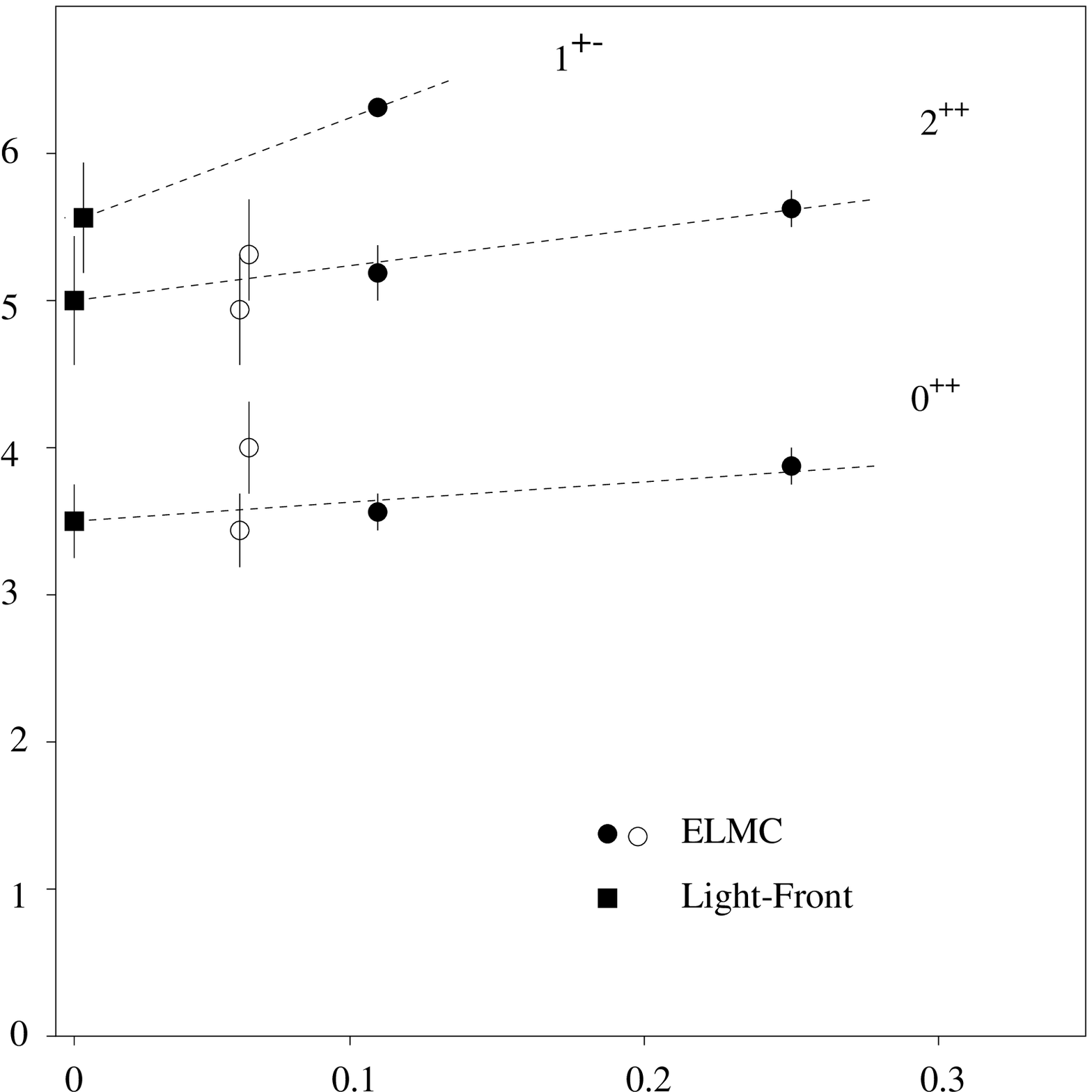 scaled 600}\\
\hspace{0.5in}$\displaystyle 1/N^{2}$
\caption{The variation of glueball masses with $N$ (pure glue).  ELMC
predictions are continuum ones for $N=2,3$ \cite{old,lattice,star} and
fixed lattice spacing estimates for $N=4$ \protect\cite{teper2}.
The dotted lines are to guide the eye and correspond to leading linear
dependence on $1/N^2$.
\label{alln}}
\end{figure}

We end this introduction by quoting the most readily understandable of
our main results: the lowest glueball masses.  We estimate the
lightest glueball in $SU(\infty)$ gauge theory, the ${\cal J^{PC}} =
O^{++}$, to be at a mass $M_{O^{++}} = 3.50 \pm 0.24 \sqrt{\sigma}$,
where $\sigma$ is the string tension. This is essentially
indistinguishable from the result for $SU(3)$ pure gauge theory
established rigorously 
by ELMC in recent years \cite{lattice}.  For the $2^{++}$
tensor state, not all components of the multiplet are yet behaving
covariantly; but from those components which are reliable, we estimate
$M_{2^{++}} = 4.97 \pm 0.43 \sqrt{\sigma}$.  The vector $1^{+-}$ has
$M_{1^{+-}} = 5.57 \pm 0.4 \sqrt{\sigma}$. We find no light
pseudo-scalar $0^{-+}$, but suspect our candidate for this has a large
error. These results are summarised in Figure~\ref{alln}.  Other
recent estimates of glueball masses have been obtained in $SU(3)$
Hamiltonian gauge theory \cite{ham} and based on extensions of the
Maldacena conjecture for large-$N$ gauge theory \cite{gravity}, though
we consider those results to be less rigorous than ours.

\section{Transverse lattice Hamiltonians.}
\label{construct}

\subsection{Energy-momentum}

In $3+1$ spacetime dimensions we introduce a square lattice of spacing $a$
in the `transverse' directions ${\bf x}=\{x^1,x^2\}$ 
and a continuum in the $\{x^0,x^3\}$ directions. In light-front (LF)
co-ordinates $x^{\pm} = (x^0 \pm x^3)/\sqrt{2}$, we treat $x^+$
as canonical time and place anti-periodic boundary conditions
on $x^- \sim x^- + {\cal L}$. Both $1/a$ and ${\cal L}$ are
high-energy cut-offs for the LF Hamiltonian
$P^- = (P^0 - P^3)/\sqrt{2}$ that evolves the system in
LF time $x^+$. The Lorentz indices $\mu, \nu \in \{ 0,1,2,3 \}$ are
split into LF indices $\alpha,\beta \in \{+,-\}$
and transverse indices $r,s\in \{1,2\}$.

The gauge field degrees of freedom below the cut-offs are represented by
Hermitian gauge potentials $A_{\alpha}({\bf x})$ and 
complex link variables $M_r({\bf x})$. 
We also introduce heavy scalar sources $\phi({\bf x})$. On the
transverse lattice, $A_{\alpha}({\bf x})$ and $\phi({\bf x})$ are
associated with a site ${\bf x}$, while $M_r({\bf x})$ is associated
with the link from ${\bf x}$ to ${\bf x} + a \hat{r}$. Each
`site' ${\bf x}$ is in fact a two-dimensional plane spanned by 
$\{x^+, x^-\}$.  These variables transform under transverse lattice gauge
transformations $U \in SU(N)$ as
\begin{eqnarray}
        A_{\alpha}({\bf x}) & \to & U({\bf x}) A_{\alpha}({\bf x}) 
        U^{\da}({\bf x}) + {\rm i} \left(\partial_{\alpha} U({\bf x})\right) 
        U^{\da}({\bf x})  \nonumber \\
        M_r({\bf x}) &  \to & U({\bf x}) M_r({\bf x})  
        U^{\da}({\bf x} + a\hat{r})  \\
        \phi({\bf x}) & \to & U({\bf x}) \phi({\bf x }) \ .
\end{eqnarray}
Since it will be 
possible to eliminate $A_{\alpha}$ by partial gauge-fixing,
$M$ and $\phi$ represent the physical transverse polarisations.
The simplest gauge covariant combinations are 
$M$, $\phi$, $F^{\alpha \beta}$, ${\rm det} M$, $\D^{\alpha} M$,
$D^{\alpha} \phi$, where the  covariant derivatives are
\begin{eqnarray}
        \D_{\alpha} M_r({\bf x})  
        & =  & \left(\partial_{\alpha} +i A_{\alpha} ({\bf x})\right)
        M_r({\bf x})  
-  i M_r({\bf x})   A_{\alpha}({{\bf x}+a \hat{r}}) 
\\
D_\alpha \phi & = & \partial_\alpha \phi +i  A_\alpha \phi \ .
\label{covdiv}
\end{eqnarray}
From these we wish to construct general LF 
Hamiltonians $P^-$ invariant
under gauge transformations and those Lorentz transformations
unviolated by the cut-offs. 

To proceed, we must make some
assumptions about which finite sets of operators to
include in the calculation. Since symmetries will be tested
explicitly, a poor choice of operators would show up 
later on. 
The following criteria were used to select operators in $P^-$ for pure
gauge theory:
%
%
\newcommand{\ci}{{(A)}}
\newcommand{\cii}{{(B)}}
\newcommand{\ciii}{{(C)}}
\newcommand{\civ}{{(D)}}
%
%
\begin{description}
\item[\ci] Canonical quadratic form for $P^+$; 
\item[\cii] Na\"{\i}ve parity restoration as ${\cal L} \to \infty$;
\item[\ciii] Transverse locality;
\item[\civ] Expansion in gauge-invariant powers of $M$.  
\end{description}
Each of these criteria deserves some explanation. The last three can
all be straightforwardly checked in principle by systematically
relaxing the condition.
\begin{description}

\item[\ci] Of the generators of the Poincar\'e group $\{P^+, P^-, P^r,
M^{\mu \nu} \}$, the subset $\{ P^r, P^+, M^{+r}$, $M^{12}$,
$M^{+-}\}$ can usually be made kinematic. That is, they can be made
independent of interactions, quadratic in the fields.  The
imposition of a lattice cut-off can spoil this property, especially if
one wants to maintain gauge invariance.  We try to maintain as many
operators as possible in kinematic form, consistent with gauge
invariance.  Further details are given below, but for now we note that
condition \ci\ can be satisfied in light-front gauge $A_{-}=0$ by
using a Lagrangian containing the only $x^+$-dependent gauge-covariant
term quadratic in link fields, $\D_{\alpha} M_r({\bf x}) (\D^{\alpha}
M_r({\bf x}))^\da$.  Although higher order terms in $M$ cannot be
ruled out, only with a quadratic term is quantisation straightforward
and, even then, only in the LF gauge $A_{-} = 0$.

\item[\cii] We will extrapolate to the ${\cal L} \to
\infty$ limit in the longitudinal direction, deriving $P^-$ from a
Lagrangian including only dimension 2 operators with respect to $\{x^+
, x^- \}$ co-ordinates. It has been noted that functions of
dimensionless ratios of longitudinal momenta $p^+$ can appear in couplings of
boost-invariant LF Hamiltonians \cite{wilson1}.
However, these functions
must be strongly constrained by LF parity $x^+ \leftrightarrow
x^-$. Parity is a dynamical symmetry on the light-front (it does not
preserve the quantisation surface) and is difficult to check
explicitly.  We will assume\footnote{This assumption is not warrented in the 
presence of finite-mass fermions.
The subtlety lies in $p^+ = 0$ vacuum modes that
are not recovered even as ${\cal L} \to \infty$. For Yukawa
interactions of finite-mass fermions, parity certainly isn't recovered
in the na\"{\i}ve way \cite{yuk}.} that, 
if $P^-$ is derived from a Lagrangian with na\"{\i}ve
$x^+ \leftrightarrow x^-$ symmetry ($p^+$-independent
couplings), then parity is correctly restored in
transverse lattice gauge theory 
in the limit ${\cal L} \to \infty$.  
Although DLCQ treatments of analogous
$1+1$-dimensional gauge theories have been successful under similar
assumptions \cite{igor,fran1,brett}, in the present case there are
many more possible operators that are ruled out by conditions \ci\ and
\cii\ combined.

\item[\ciii] In products of gauge-invariant operators on the
transverse lattice, each term of the product can be arbitrarily
separated in the transverse direction. We assume some transverse
locality by restricting products of gauge invariant operators to share
at least one site ${\bf x}$.
 
\item[\civ] After fixing to LF gauge $A_{-} = 0$ and eliminating the
resultant constrained field $A_{+}$, Fock space states will consist of
link-partons derived from the Fourier expansion of $M$.  For
sufficiently large $a$, $M$ remains a massive degree of freedom and
there is an energy barrier for the addition of a link-parton to a Fock
state.  Operators of order $M^p$ in $P^-$ will connect Fock states
differing by at most $p-2$ link-partons. By expanding $P^-$ to a given
order of $M$ in this regime (the colour-dielectric expansion)  we
therefore cut off interactions between lower-energy few-parton states
and higher-energy many-parton states. Note that all energy scales are
still allowed however, since we take the ${\cal L} \to \infty$ limit,
enabling highly virtual small $p^+$ partons to appear. The advantage
of a cut-off on changes of parton number is that it organises states
into a constituent hierarchy, consistent with energetics.

\end{description}

The Lagrangian density satisfying condition \ci\ can be written
\be 
L_{\bf x} = \D_{\alpha} M_r({\bf x}) (\D^{\alpha} M_r({\bf x}))^\da 
- V_{\bf x} -U_{\bf x}
\eq
where the `potential' has a purely transverse part $V_{\bf x}$ and a
mixed part $U_{\bf x}$.  Up to 4th order in $M$, the purely transverse
part is
\begin{eqnarray}
 V_{\bf x} & = & 
-{\beta \over Na^{2}} \Tr\left\{ M_{1} ({\bf x}) 
M_{2} ({\bf x} + a \hat{1})
M_{1}^{\da}
({\bf x} + a \hat{2} )  
M_{2}^{\da}({\bf x})
\right\} + \ {\rm c.c.} \nonumber \\
\nonumber \\
&& +  \sum_r  
 \mu^2  \Tr\left\{M_r M_r^{\da}\right\} 
+ \sum_r {\lambda_0 \over a^{2} N} \left( \det [M_{r}]  + \det
[M_{r}^{\da}] \right) 
\nonumber\\
&& + \sum_r  {\lambda_1 \over a^{2} N}
\Tr\left\{ M_r M_r^{\da}
M_r M_r^{\da} \right\} 
+
 {\lambda_2 \over a^{2} N}\sum_r  
\Tr\left\{ M_r ({\bf x}) M_r({\bf x} + a \hat{r} )
M_r^{\da}({\bf x} + a \hat{r} ) M_r^{\da} ({\bf x})\right\} \nonumber \\
&& + \sum_r   {\lambda_3 \over a^{2} N^2} 
\left( \Tr\left\{ M_r M_r^{\da} \right\} \right)^2
+  {\lambda_4 \over a^{2} N}  
\sum_{\sigma=\pm 2, \sigma^\prime = \pm 1}
        \Tr\left\{ 
M_\sigma^{\da} M_\sigma M_{\sigma^\prime}^{\da} M_{\sigma^\prime} \right\} 
        \nonumber\\
&& +  {4 \lambda_5 \over a^{2} N^2} 
\Tr\left\{ M_1 M_1^{\da} \right\}\Tr\left\{ M_2 M_2^{\da} \right\} 
\; . \label{pot}
\end{eqnarray}
The $\det [M]$ term can be dropped if $N>4$
and can always be neglected in the large-$N$ limit.
The mixed part, to the approximation we will need, can be written
\be
U_{\bf x}  = \sum_{\bf y} \epsilon_{\alpha \beta}  \Tr \ 
\{ E_{\bf x} P_{\bf xy} F^{\alpha \beta}({\bf y})
P_{\bf xy}^{\da} \}  - G^2 f(E_{\bf x}) \ , \label{gen}
\eq
where $E_{\bf x}$ is a non-dynamical pseudoscalar adjoint
field at site ${\bf x}$, while $P_{\bf xy}$ is a linear combination of
Wilson lines in $M$, each from  from ${\bf x}$ to ${\bf y}$ (for
gauge invariance).
Upon integrating out $E$, the function $f(E)= \Tr\left\{E^2\right\} + O(E^4)$ 
gives a  simple 
${1 \over 2 G^2} \Tr \left\{ F^{\alpha \beta} F_{\alpha \beta} \right\}$ 
term if only the $E^2$ 
part is retained and we set
$P_{\bf xy} = \delta_{\bf xy}$.\footnote{The case $P_{\bf xy} =
\delta_{\bf xy}$ is related to  the generalised 2DQCD of
M. Douglas {\em et al.}\ \cite{genqcd}} This will be the only
term needed for pure gauge theory to $O(M^4)$ once $A_{\alpha}$ has
been eliminated. We gave the more general form in Eqn.~(\ref{gen}) since
it will be relevant in the heavy-source analysis.

From the above Lagrangian, four 
of the usual seven kinematic Poincar\'e
generators derived canonically from the energy-momentum tensor $T^{\mu
\nu}$ remain gauge invariant and can be made kinematic by
LF gauge choice $A_{-}=0$. They are, at $x^+ = 0$ say,
\begin{eqnarray}
P^+ & = & 2 \int dx^- \sum_{{\bf x}, s} \Tr  
                        \left\{ \partial_- M_s({\bf x})  
                \partial_- M_s({\bf x})^{\da} \right\} \ , \label{mom} \\
M^{+-} & = &  2 \int dx^- \sum_{{\bf x}, s}  x^- 
             \Tr\left\{ \partial_{-} M_s({\bf x}) 
             \partial_{-} M_s({\bf x})^{\da} \right\}  \\
M^{+r} & = & - 2 \int dx^- \sum_{{\bf x}, s} 
      \left( x^r + \frac{a}{2} \delta^{rs}\right)
                      \Tr  \left\{ \partial_- M_s({\bf x})  
                \partial_- M_s({\bf x})^{\da} \right\} 
\end{eqnarray}
We will use these to define states of definite momentum $\{P^+,
{\bf P} \}$. The other three would-be kinematic Poincar\'e generators 
in light-front formalism, $\{ P^r, M^{12} \}$, are not gauge-invariant
when derived canonically from $T^{\mu \nu}$ because of the lattice
cut-off; the simplest gauge-invariant extensions are no longer 
quadratic in fields, even in LF gauge. 
 
Of the dynamic generators, the most important, and the only one
we explicitly treat, is the light-front Hamiltonian itself
\begin{eqnarray}
P^-  & = & \int dx^- \sum_{{\bf x}}  \left(
                V_{{\bf x}} - \Tr  \left\{ A_{+}({\bf x})
        J^{+}({\bf x}) \right\}  
- {1 \over G^2}
 \Tr \left\{ \partial_{-}A_{+}\partial_{-}A_{+} \right\}\right) \; , \\
J^{+}({\bf x}) &=&  i \sum_{r} \left(
M_r ({\bf x}) \stackrel{\leftrightarrow}{\partial}_{-} 
M_r^{\da}({\bf x})  + M_r^{\da}({\bf x} - a\hat{r}) 
\stackrel{\leftrightarrow}{\partial}_{-} M_r({\bf x} - a\hat{r})
\right) \; .
\end{eqnarray}
Rather than trying to directly construct an approximate  realisation of the
Poincar\'e algebra at finite cut-off, we will minimise cut-off
artifacts by optimising restoration of Lorentz covariance in
low-energy eigenstates of the Hamiltonian $P^-$.  

In the light-front gauge, $A_+$ is a non-dynamical variable and eliminating
it introduces non-local interactions thus
\begin{eqnarray}
 P^-  &  = &  \int dx^- \sum_{{\bf x}}\left(
  {G^2 \over 4} \Tr\left\{ {J^{+} \over \partial_{-}} 
{J^{+} \over \partial_{-}} \right\} 
- {G^2 \over 4N} \Tr \left\{ {J^{+} \over \partial_{-}}   \right\}
\Tr \left\{ {J^{+} \over \partial_{-}}   \right\} + V_{{\bf x}}\right)
\end{eqnarray}
where $J^+ / \partial_{-} \equiv \partial_{-}^{-1}(J^+)$.  There is
still a residual $x^-$-independent gauge invariance generated by the
charge $\int dx^- J^+$. As originally shown in
Refs.~\cite{bard1,bard2}, finite energy states $|\Psi\rangle$ are
subject to the gauge singlet condition $\int dx^- J^+ |\Psi\rangle =
0$.  In the large-$N$ limit, this means that Fock space at fixed $x^+$
is formed by connected closed loops of link variables $M$ on the
transverse lattice (the $x^-$ co-ordinate of each $M$ is
unrestricted).

\subsection{Quantisation}

The dynamical problem is now to diagonalise $P^-$ at fixed total
momenta $\{P^+, {\bf P} \}$. A convenient basis consists of
free link-partons obtained from the Fourier modes $a(k^+,{\bf x})$
of $M$ in the $x^-$ co-ordinate
\begin{eqnarray}
M_r(x^+=0,x^-, {\bf x}) &  =  & 
        \frac{1}{\sqrt{4 \pi }} \int_{0}^{\infty} {dk^+ \over {\sqrt{ k^+}}}
        \left( a_{-r}(k^+, {\bf x})\, e^{ -i k^+ x^-}  
+   a^{\da}_r(k^+, {\bf x})\, e^{ i k^+ x^-} \right) \ .
\end{eqnarray}
The Fock space formed from creation operators $a^{\da}$ in the
large-$N$ limit, and other details of the calculation in this Fock space,
including construction of states of definite momentum,
have been described elsewhere \cite{dv1,dv2}. 
We applied both DLCQ and Tamm-Dancoff cut-offs in Fock space,
extrapolating both of these at fixed values of the couplings
in Eqn.~(\ref{pot}).\footnote{The largest cut-offs used were
a $K=26/2$ DLCQ resolution and 8-link Tamm-Dancoff cut-off in the
glueball sector, while a $K=70/2$ and 4-link cut-off was used in the
heavy-source sector. The latter sector has an additional cut-off
$P^+_{\rm max}$, as discussed in Ref.~\cite{dv2}, which was varied up to 7.} 

Low-energy eigenfunctions of $P^-$ ({\em id est} glueballs) are to be
tested for Lorentz covariance, the couplings appearing in Eqn.~(\ref{pot})
being tuned to minimise covariance violations.  Since $G^2 N$, with
dimension $\left(\mbox{energy}\right)^2$,
is consistent with 't Hooft's large-$N$
limit \cite{hoof}, we will use it to set the dimensionful scale.
Thus, the dispersion
relation of a glueball can be written
\begin{eqnarray}
      2P^+ P^- & = &  G^2 N \left( \dm^{2}_{0} + \dm_{1}^{2}\, a^2
      |{\bf P}|^2  
+ 2 \overline{\dm}_{1}^{2} \, a^2 P^1 P^2  
+ O(a^4 |{\bf P}|^4) \right) \label{latshell}\; .
\end{eqnarray}
For each glueball, $\dm^{2}_{i}$ are dimensionless functions of the
couplings which, for a given set of couplings, are extracted by
expanding eigenvalues of $P^-$ in $a {\bf P}$.  A truly relativistic
state must have an isotropic speed of light
\begin{eqnarray}
a^2 G^2 N \dm_{1}^{2}  \equiv c_{\rm on}^2 & = & 1 \nonumber \\
a^2 G^2 N (\dm_{1}^{2}+\overline{\dm}_{1}^{2}) \equiv 
c_{\rm off}^2 & = & 1  \ . 
\end{eqnarray}
$c_{\rm on}$ is the speed of light in direction ${\bf x}=(1,0)$;
$c_{\rm off}$ is the speed of light in the direction ${\bf x}=(1,1)$.
In addition, the an-harmonic corrections at $O(a^4 |{\bf P}|^4)$ must
vanish, but we will not use this condition directly.

The ratio of transverse to longitudinal scales is set by the
dimensionless combination $a^2 G^2 N$. This can be deduced from two
measurements of the string tension $\sigma$, for example
\cite{burk}.  The first
measures the mass squared of winding modes on a periodic transverse
lattice, fitting the groundstate to $n^2 a^2 \sigma^{2}_{T}$ for
winding number $n$.  The second measures the heavy-source potential in
the continuous $x^3$ direction, fitting the groundstate to
$\sigma_L R$ for separation $R$.\footnote{One could also use the
heavy-source potential in transverse directions to find $\sigma_{T}$,
but it is more accurate to use winding modes for this.}  Demanding
$\sigma_T=\sigma_L \equiv \sigma$ and expressing eigenvalues in units
of $G^2 N$, we may eliminate $\sigma$ and deduce $a^2 G^2 N$. We may
also deduce $G^2 N$, and hence all dimensionful quantities, in units
of $\sigma$. Some refinements we have made to the calculation of the
heavy-source potential, since our previous work, are described in the
next subsection.

\subsection{Heavy-source potential $V_{QQ}$}
\label{heavy}

The non-relativistic
heavy-source potential $V_{QQ}$ provides us with a number of important
pieces of information. The string tension
$\sigma$ will be used as our basic physical unit for dimensionful
quantities. The deviations of the potential from rotational invariance
will help to fix couplings in the theory. We find that, in general, both
relativistic invariance in the glueball sector and rotational invariance
in the heavy-source sector are needed to accurately
pin down the  effective
LF Hamiltonian.

The heavy-source sector was first studied on the transverse lattice in
$2+1$ dimensions by Burkardt and Klindworth \cite{burk}. They used the
most rudimentary approximation to the Lagrangian, quadratic in the
link and heavy-source fields.  Along both the continuous and lattice
spatial directions linear potentials arise, which can be given the
same slope $\sigma$ by adjusting the link-field mass $\mu^2$ for given
$a/\sqrt{\sigma}$.  The authors of Ref.~\cite{burk} observed that the
contours of the potential in the spatial plane were `square' at $\mu^2
=\infty$ ($a\sim \infty$) and `circular' at $\mu^2 =0$ ($a \sim
1/\sqrt{\sigma}$).  This rounding of the potential indicates that
rotational invariance is improved as the lattice spacing is reduced,
as one would hope. However, there is still a potentially large
violation of rotational invariance due to the fact that any
discrepancy between the tensions measured in continuum and lattice
directions can be masked by adjusting $a$. Only by examining another
sector where the lattice spacing value is used, such as covariance of
glueballs, can one tell whether the potential is truly circular or
merely oval. Using more accurate higher order LF Hamiltonians in the
pure glue sector \cite{dv2,dv3}, we found that the values of the
lattice spacing deduced from rotational invariance of the heavy-source
potential were not very compatible with those needed for relativistic
covariance of glueballs. Therefore one needs to also improve the
accuracy in the heavy-source sector. We take some steps in this direction
in the following.

Our light-front treatment of heavy-source fields $\phi_i$, which we
take to be scalars in the fundamental representation, has been
introduced in Ref.~\cite{dv2}. We will work with a LF Hamiltonian in
LF gauge $A_- = 0$ to order $M^4$ in the pure glue operators, order
$M^2$ in the heavy-source operators, leading order of the heavy-source
mass $\rho$, together with the conditions discussed in
Section~\ref{construct}.  This LF Hamiltonian for heavy sources with
non-zero spatial separation can be derived from the Lagrangian
density
\begin{eqnarray}
L_{\bf x}  & = & \sum_{r}
  \Tr\left\{\D_{\alpha} M_r({\bf x})\, 
       \left(\D^{\alpha} M_r({\bf x})\right)^\da\right\} 
- \frac{1}{2 G^2} \Tr \left\{ F^{\alpha \beta} F_{\alpha \beta} \right\} 
-  V_{\bf x} 
   +  \left(D_\alpha \phi\right)^\da D^\alpha \phi 
        - \rho^2 \phi^\da \phi 
\nonumber \\ && 
- \frac{\tau_1}{N G^2}
\Tr \left\{ F^{\alpha \beta}\, F_{\alpha \beta}\, W_r \right\} 
 - \frac{\tau_2}{N G^2}
\Tr \left\{ M_{r}^{\da}({\bf x})\, F^{\alpha \beta}({\bf x})\, M_{r} ({\bf x})
 \, F_{\alpha \beta}({\bf x} +
a \hat{r}) \right\} \ , \label{qqlag}
\end{eqnarray}
where
\be
W_r = \left(M_{r}^{\da} M_{r}+M_{r} M_{r}^{\da}
        \right) \ .
\eq
In addition, there are combinations which are 
suppressed in
the $\rho \to \infty$ limit, such as $\phi^\da W_r \phi$, $\phi^\da
F^{\alpha \beta} F_{\alpha \beta} \phi$, which we did not include in
this work.\footnote{We investigated the first of these terms in 
earlier work by scaling its
coefficient with $\rho$ so that it survived the $\rho \to \infty$
limit. It had 
little affect on the string tension but seemed to help short-distance
rotational invariance.}

After gauge fixing $A_{-}=0$, eliminating $A_{+}$ in powers of
$M$, and discarding the higher orders in $M$,
the LF Hamiltonian resulting from (\ref{qqlag}) is
\begin{eqnarray}
 P^-  &  = &  \int dx^- \sum_{{\bf x}, r}
 {G^2 \over 4} \Tr\left\{ {J^{+}_{\rm tot} \over \partial_{-}}
{J^{+}_{\rm tot} \over \partial_{-}} 
\right\} 
- {G^2 \over 4N} \Tr \left\{ {J^{+}_{\rm tot} \over \partial_{-}} \right\}
\Tr \left\{ {J^{+}_{\rm tot} \over \partial_{-}} \right\}
+ V_{{\bf x}} \nonumber \\ && 
+ \rho^2 \phi^\da \phi 
+ {2 \tau_1 \over N} \Tr\left\{ {J^{+} \over \partial_{-}}
{J^{+} \over \partial_{-}} W_r 
\right\} 
+ { 2 \tau_2 \over N} \Tr\left\{ {J^{+}({\bf x}) \over \partial_{-}}
M_{r}({\bf x}) {J^{+}({\bf x} + a \hat{r}) \over \partial_{-}}
M_{r}^{\da}({\bf x})
\right\} 
\end{eqnarray}
with 
\be
    J^{+}_{\rm tot}  = J^+ + i \phi \stackrel{\leftrightarrow}{\partial}_{-}
 \phi^\da 
\eq
The diagonalisation of this Hamiltonian in Fock space proceeds in a
similar way to the glueball sector (see Ref.~\cite{dv2}).

The selection (\ref{qqlag}) is  more general than was used
in previous calculations.
Physically speaking, the new terms $\tau_1$ and
$\tau_2$ generate oscillations of the flux string perpendicular to
$x^3$-separated sources (via enhanced pair production of links) as
required of a rough gauge string \cite{rough}. This partially screens
the linear potential in the $x^3$-direction to produce a rotationally
invariant $\sigma$. However, while we find that these new terms
improve the interaction between heavy sources at zero and one-link
transverse separation, small-$x^3$ separations and wider transverse
separations are left unimproved. This will result in a loss of
off-axis rotational invariance, which could only be remedied by
including terms at $O(M^4)$ {\em et cetera}. The underlying difficulty
in getting rotational invariance may be related to the existence of a
roughening transition separating the large $a$ regime from the
continuum in the heavy-source sector.

\section{Space of Hamiltonians}
\label{couplings}

Our procedure for extracting Lorentz-covariant cut-off independent
results is non-standard, so we take some time in this section to
describe the procedure we used for the present calculation.

\subsection{Topology}

Abstractly, we have a set of cut-off Hamiltonians
containing
in principle an infinite number of operators $O_i$ with couplings
$g_i$. For simplicity, we consider only one high-energy cut-off $1/a$,
all others having been removed by extrapolation.\footnote{This 
extrapolation in
itself may require tuning of the $g_i$ in general. However, such
tuning is not necessary for pure gauge theories within the
approximations we make.}  ${\cal H}$ is the infinite-dimensional space
of the $g_i$, the space of Hamiltonians.

One could in ${\cal H}$ define renormalisation group trajectories as
trajectories along which, say, the eigenvalues of the Hamiltonian 
are invariant.
We are particularly interested in a trajectory on which Lorentz
covariance is restored ---  the `Lorentz trajectory'
${\cal T}$ --- for observables containing no explicit momenta above
the cut-off.  Perry and Wilson \cite{perry} have emphasised the idea, albeit in
weak-coupling perturbation theory, that when a symmetry is restored in
a cut-off theory, the number of couplings running independently with
the cut-off is reduced.  For Lorentz symmetry in pure
gauge theory, for example, this coherent running of the couplings
should pick out ${\cal T} \subset {\cal H}$. A
basic hypothesis of our approach to pure $SU(N)$ gauge theory on the
transverse lattice will be that a single one-parameter Lorentz
trajectory extends from the continuum to quite low cut-offs.  In order
to test the hypothesis, we will specify conditions for an approximate
Lorentz trajectory ${\cal T}_s$ in a finite-dimensional subspace
${\cal H}_{s} \subset {\cal H}$.  The subspace ${\cal H}_{s}$ arises
from the conditions we placed on the Hamiltonian in
Section~\ref{construct}. The most straightforward way to define ${\cal
T}_s$ is in terms of a finite selection of observables contributing to
a $\chi^2$ test for deviations from Lorentz covariance.  At the same
time, we believe that it should be possible to develop tests of
Lorentz covariance that do not depend on a specific choice of
observables, {\em exempli gratia} ensuring closure of the Poincar\'e
algebra. But we have yet to carry out this development.

\begin{figure}
\centering
\BoxedEPSF{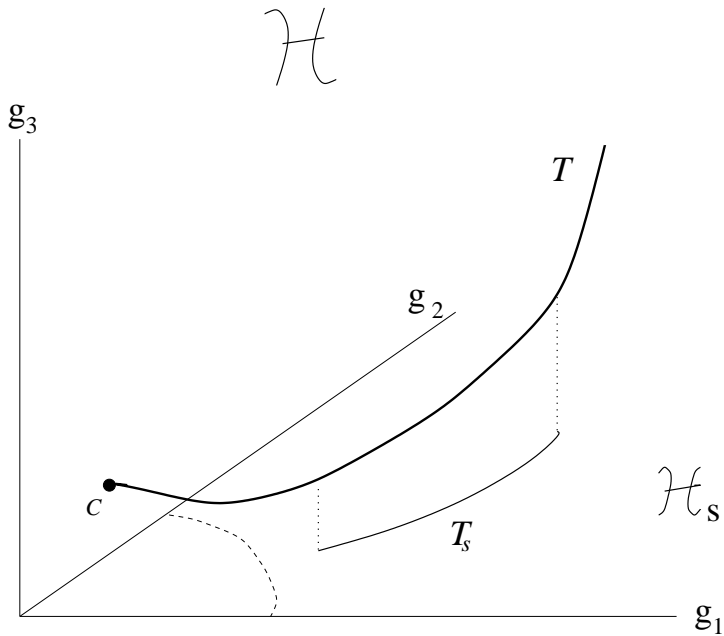 scaled 500}
\caption{The space ${\cal H}=\{ g_1, g_2 , g_3, \cdots \}$ of 
Hamiltonians and 
its subspace ${\cal H}_{s}=\{ g_1, g_2 \}$.
${\cal T}$ is the exact Lorentz trajectory, 
${\cal T}_s$ the approximation to it in
${\cal H}_{s}$. ${\cal C}$ is the intersection of ${\cal T}$ with the
continuum $1/a = \infty$. The dashed line is a possible
phase boundary in ${\cal H}_{s}$ separating 
${\cal T}_s$ from the continuum.
\label{top}}
\end{figure}
 
Figure~\ref{top} illustrates the space of Hamiltonians we have in
mind.  The space ${\cal H} = \{g_1,g_2,g_3, \cdots \}$ represents the
infinite-dimensional space of couplings. The cut-off $1/a$ decreases
along the Lorentz trajectory ${\cal T}$ as one moves away from the
intersection of ${\cal T}$ with the continuum, ${\cal C}$.  A
finite-dimensional subspace ${\cal H}_s$ is denoted by the $g_3 = 0$
plane. ${\cal T}$ will not lie entirely in ${\cal H}_s$ in general, so
the best one can do is investigate approximations in ${\cal H}_s$.  We
search ${\cal H}_s$ over a certain range of cutoffs for a trajectory
${\cal T}_s$ which minimises the distance to ${\cal T}$, {\em id est}
minimises violations of Lorentz covariance.  It is not sufficient to
simply find an isolated point in ${\cal H}_s$ at which violations of
Lorentz covariance are minimised.  This may have little to do with
${\cal T}$. One must establish an entire trajectory.  If ${\cal T}$
does not exist for the range of cut-off investigated, or lies far from
${\cal H}_s$, one will be unable to find an unambiguous trajectory
${\cal T}_s$, and the method fails.

The metric to determine the distance from ${\cal T}$ is the 
$\chi^2$ test for Lorentz covariance, which involves both a choice of
observables and their relative weights. Even if ${\cal T}$ exists,
a random choice of $\chi^2$ observables and their
weights will make it difficult to 
identify. Consistent with the colour-dielectric expansion, we
gave higher weights to low energy observables dominated by
a few link-partons and made sure
that the number of observables was always $\gg  \dim\left[{\cal
H}_s\right]$. 

Having convinced oneself that ${\cal T}_s$ is uniquely established
for the metric employed and the range of cut-off investigated, one can
use it to estimate the value of observables on ${\cal T}$.
Observables will exhibit approximate
scaling along ${\cal T}_s$, by virtue of its nearness to ${\cal T}$
(which is an exact scaling trajectory). The values of these
observables will be close to the values on ${\cal T}$, and the
violations of scaling {\em and} covariance as one moves along ${\cal
T}_s$ can be used to estimate the systematic error from use of a
finite-dimensional space of Hamiltonians.  As one enlarges the space
${\cal H}_s$, {\em exempli gratia} by going to higher orders of the
colour-dielectric expansion, one can get closer to ${\cal T}$ and the
errors reduce.

This procedure is to be contrasted with the traditional perturbative
`improvement' program of introducing irrelevant operators in a cut-off
theory.  There, one derives the couplings via perturbative
renormalisation group transformations, solves the theory, and
extrapolates to the continuum. This is the approach to LF Hamiltonians
being pursued by a number of authors \cite{ohio}. Instead, we use
symmetry directly to construct coupling trajectories, making
the assumption that symmetry uniquely fixes the values of observables
in QCD, once the overall scale has been specified.  In practice, we do
not construct renormalisation group trajectories and do not
extrapolate to the continuum.

We do not try to extrapolate the results to $a=0$ because in ${\cal
H}_s$ there may be barriers to the continuum ({\em exempli gratia}\/
roughening or large-$N$ transitions). Our use of disordered massive
link variables $M$, together with the concomitant expansion of $P^-$
in powers of $M$, means that the transverse lattice spacing $a$ is
quite coarse in the region of ${\cal H}$ we search.  However, by
definition, there can be no phase boundaries separating ${\cal T}
\subset {\cal H}$ from the continuum.  Provided a suitable ${\cal T}$
exists, and we can get close enough to it, the interpolation of phase
boundaries only breaks the connection between scaling violations in
each phase of ${\cal H}_s$, not the estimate of the observable itself.

\subsection{Topography}

We now  construct a set of topographical charts of ${\cal H}_s$ that
help one to decide whether a unique Lorentz trajectory
exists. 
It is 
convenient to use $G^2 N$ to set the dimensionful scale wherever
possible, and introduce dimensionless versions of the other 
couplings as the co-ordinates of ${\cal H}_s$
\begin{eqnarray}
        m^2 & = & {\mu^2 \over G^2 N}  \; , \;\;
        \newl_i = {\lambda_i \over a^{2} G^2 N}  \  , \;\;
        \newtau_i  =  \frac{\tau_i}{\sqrt{G^2 N}} \; , \; \;
         b = {\beta \over a^{2} G^2 N} \; .
\end{eqnarray}
We employed two ways to search ${\cal H}_s$.
\newcommand{\mi}{{Method~1}}
\newcommand{\mii}{{Method~2}}
\begin{description}

\item[\mi] For a given $m^2$, this minimises $\chi^2$ over all 
couplings $\{ l_1,
l_2, l_3, l_4, l_5, b \}$ to give a
one-parameter trajectory in ${\cal H}_s$.  This amounts to a direct
determination of ${\cal T}_s$. 

\item[\mii] 
For a given $m$ and $g_i \in \{ l_1,
l_2, l_3, l_4, l_5, b \}$, this minimises $\chi^2$ over all
the remaining couplings.
There is a chart for each $g_i$ which can be displayed as a contour plot with
heights $\chi^{2}_{min}$. The existence of ${\cal T}$ should show up
as a unique valley on each chart, universal with respect to small
changes in the form of the metric, the bottom of the valley coinciding
with the trajectory ${\cal T}_s$ found by \mi.

\end{description}
With \mi, in a reasonable amount of time we could extrapolate the
DLCQ and Tamm-Dancoff cut-offs before using a routine to iteratively
search for the minimum value of $\chi^2$, setting the mass scale from
first principles via string tension measurements.  Because of its more
time-consuming nature, a number of compromises were necessary to
obtain decent results by \mii: the DLCQ and Tamm-Dancoff cut-offs
were fixed;\footnote{$K=6/2$ and a 6-link truncation.} to avoid the extra
complication introduced by the heavy-source analysis, the mass scale
was fixed by setting ${\cal M}(0^{++}) = 3.5 \sqrt{\sigma}$, a result
which had been deduced from \mi {} on ${\cal T}_s$ (see later); we
used fixed values $l_3 = 100$ and $l_5 = 10$, again motivated by the
typical values on ${\cal T}_s$ from \mi.  The very large
couplings $l_3$ and $l_5$ have a special significance which is
explained in Ref.~\cite{dv2}. They can be set to virtually any very
large value for a Lorentz-covariant theory.

\begin{figure}
\centering
\BoxedEPSF{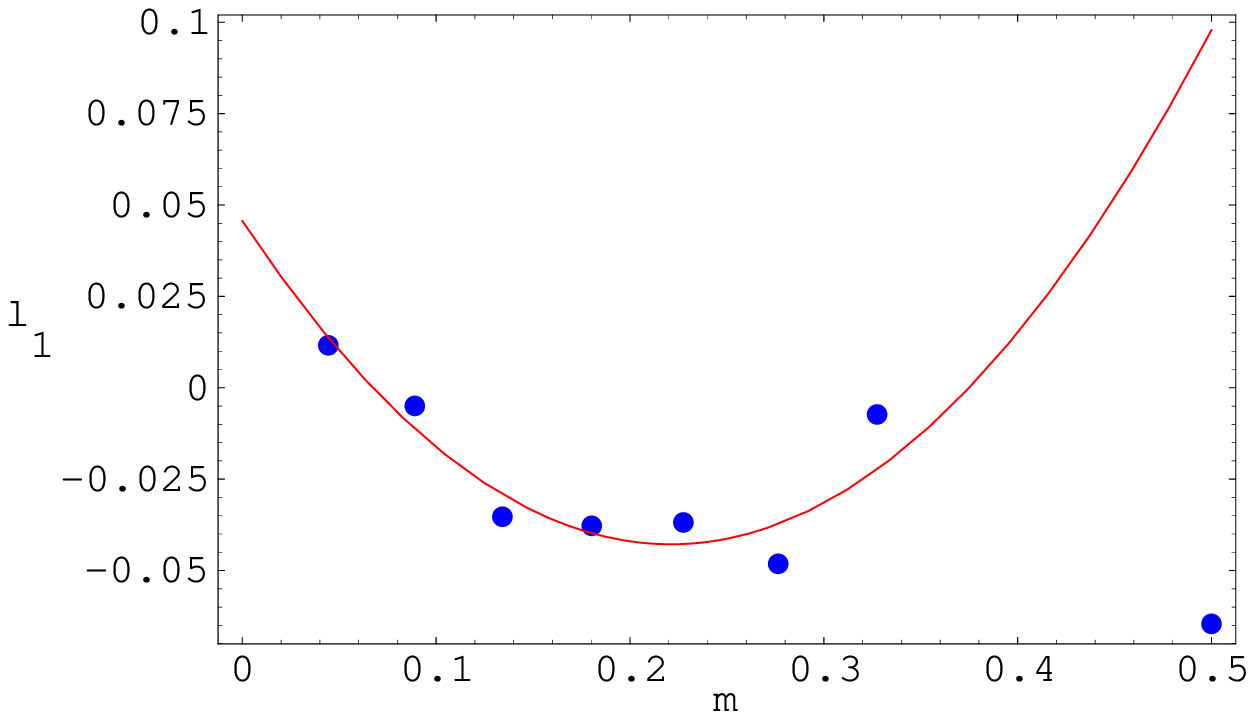 scaled 540}
\BoxedEPSF{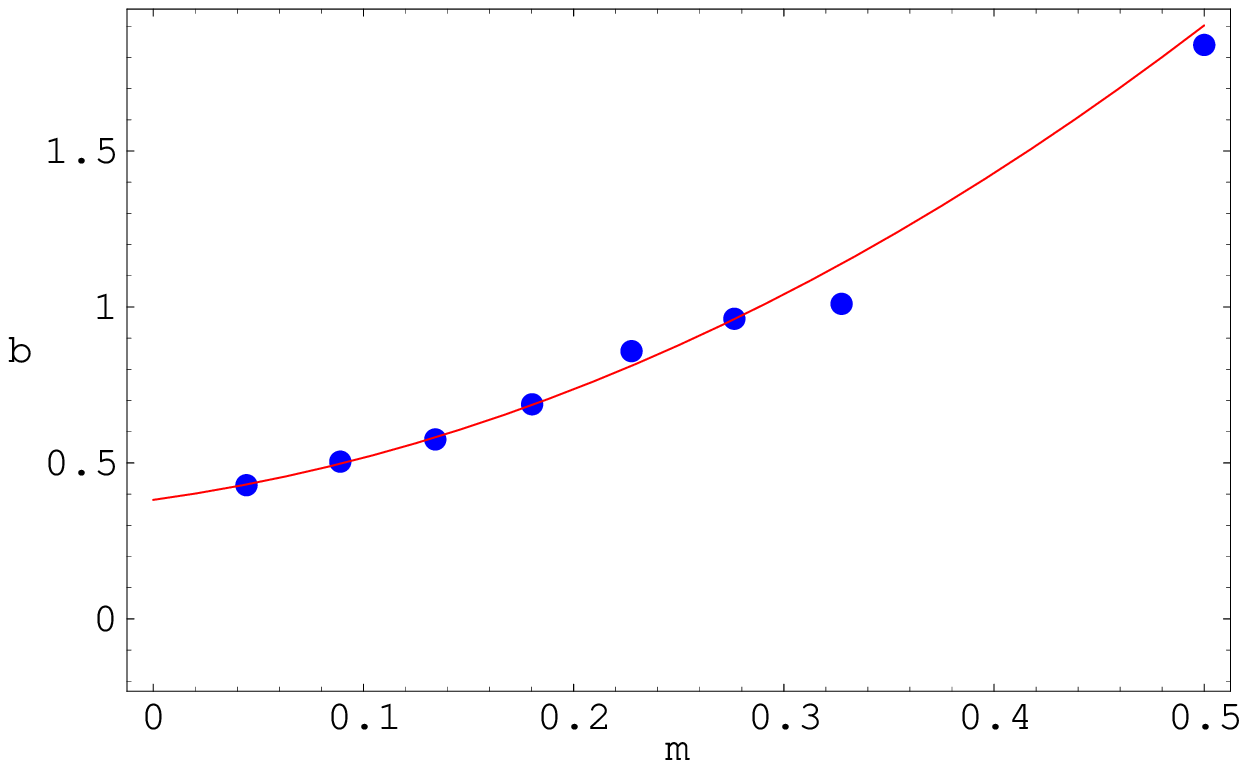 scaled 500}\\
\BoxedEPSF{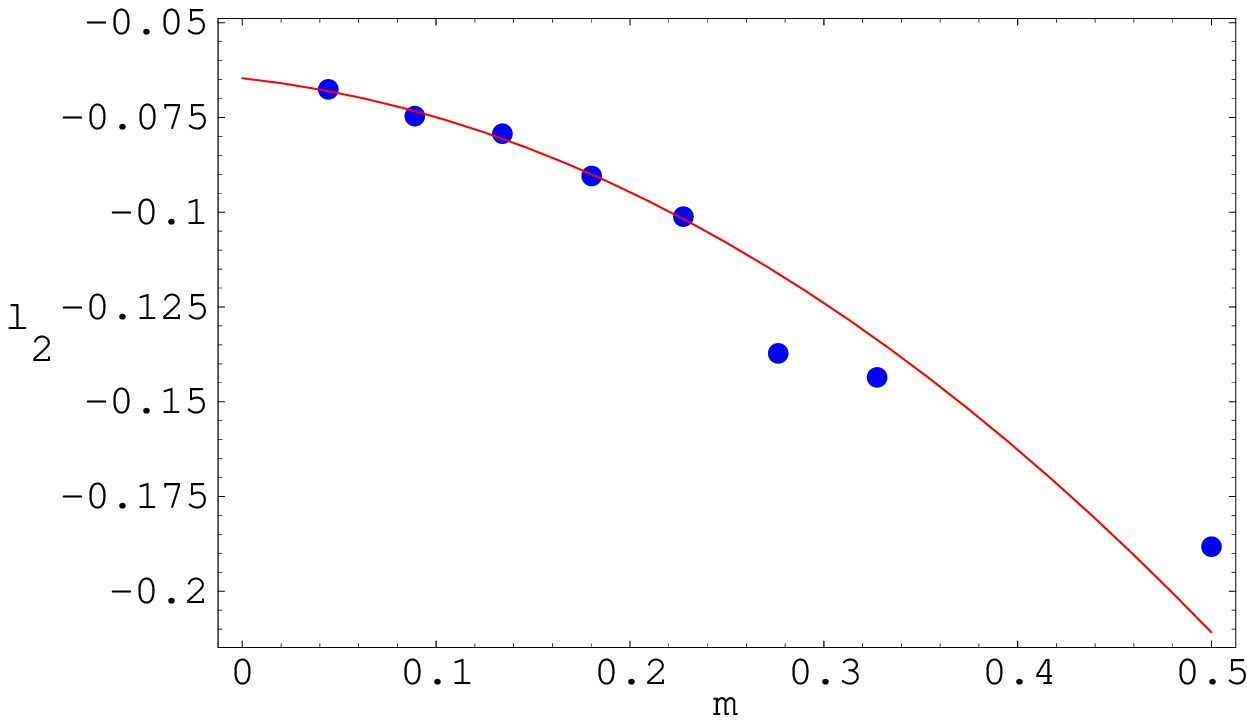 scaled 540}
\BoxedEPSF{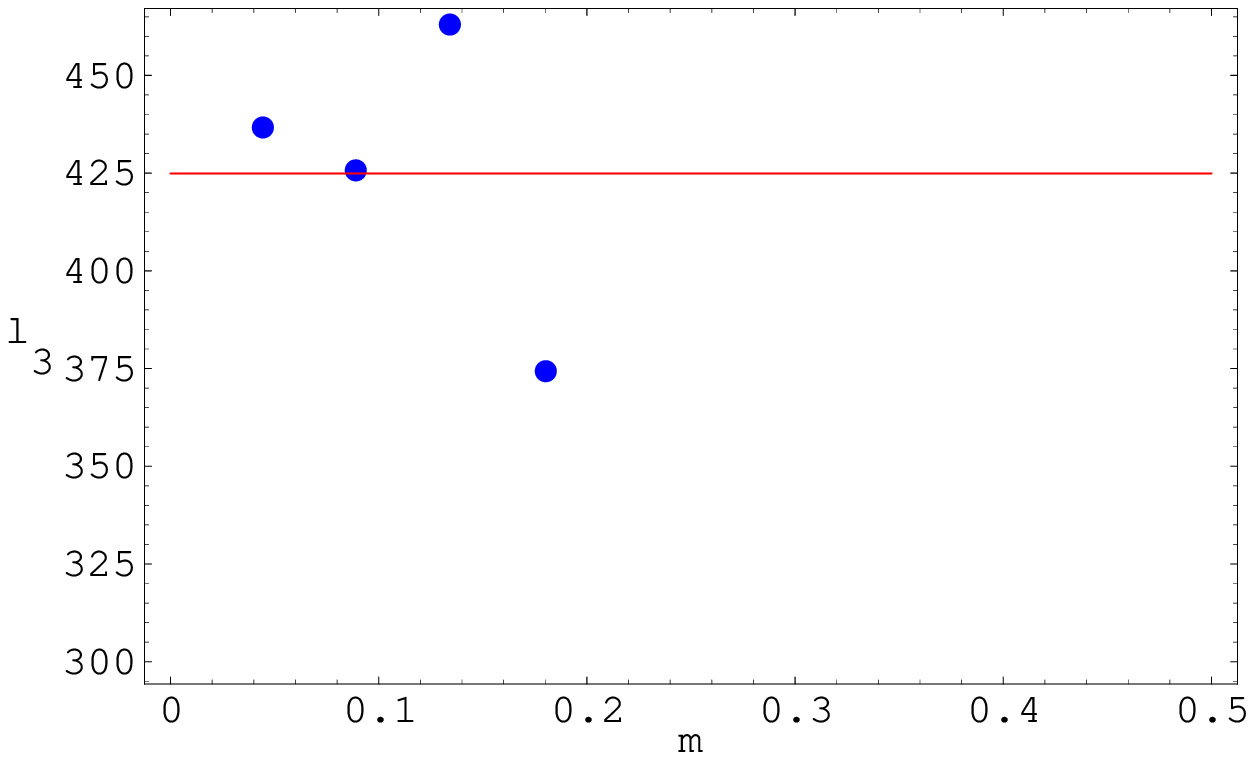 scaled 500}\\
\BoxedEPSF{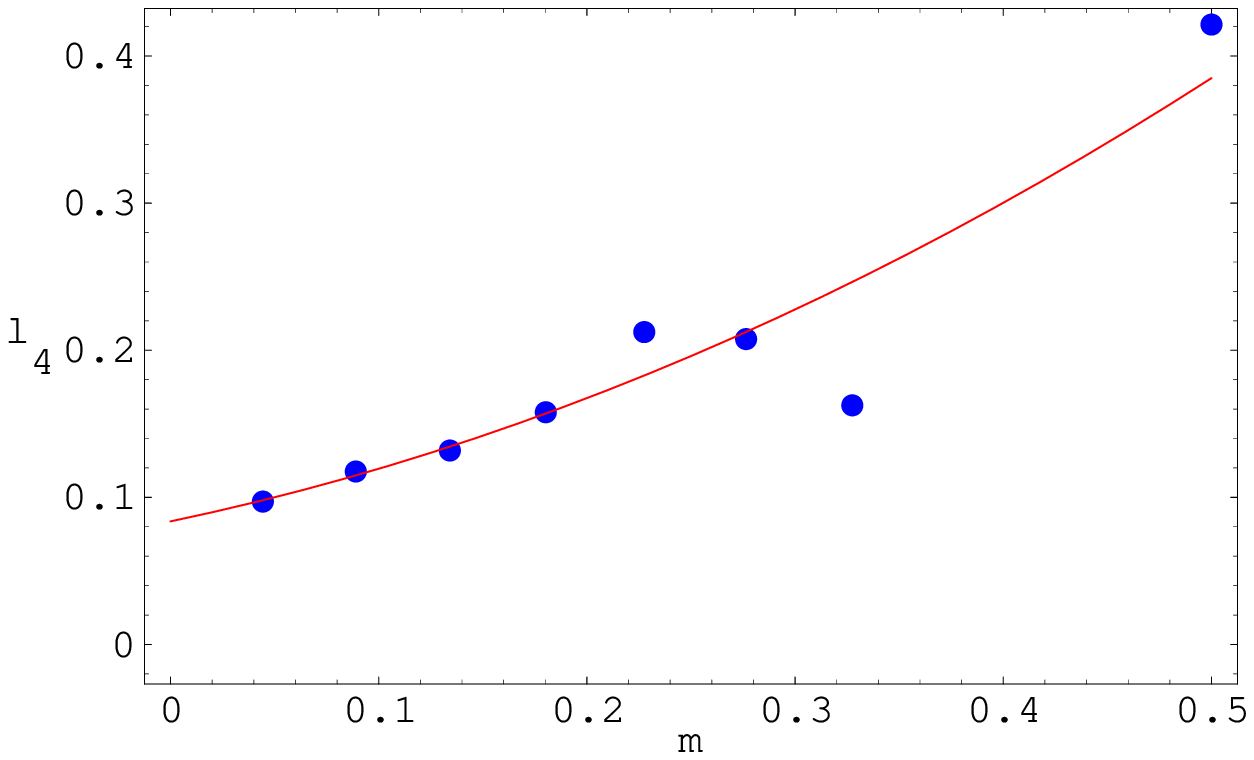 scaled 500}
\BoxedEPSF{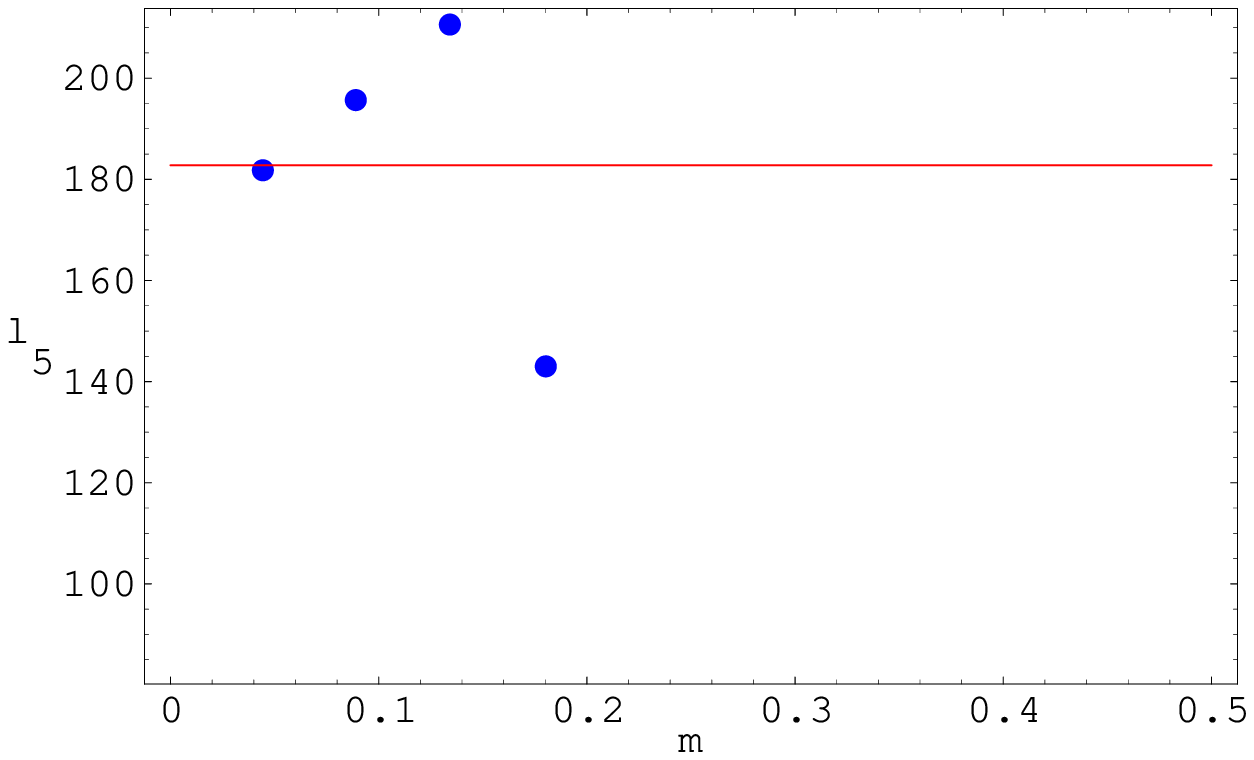 scaled 500}\\
\BoxedEPSF{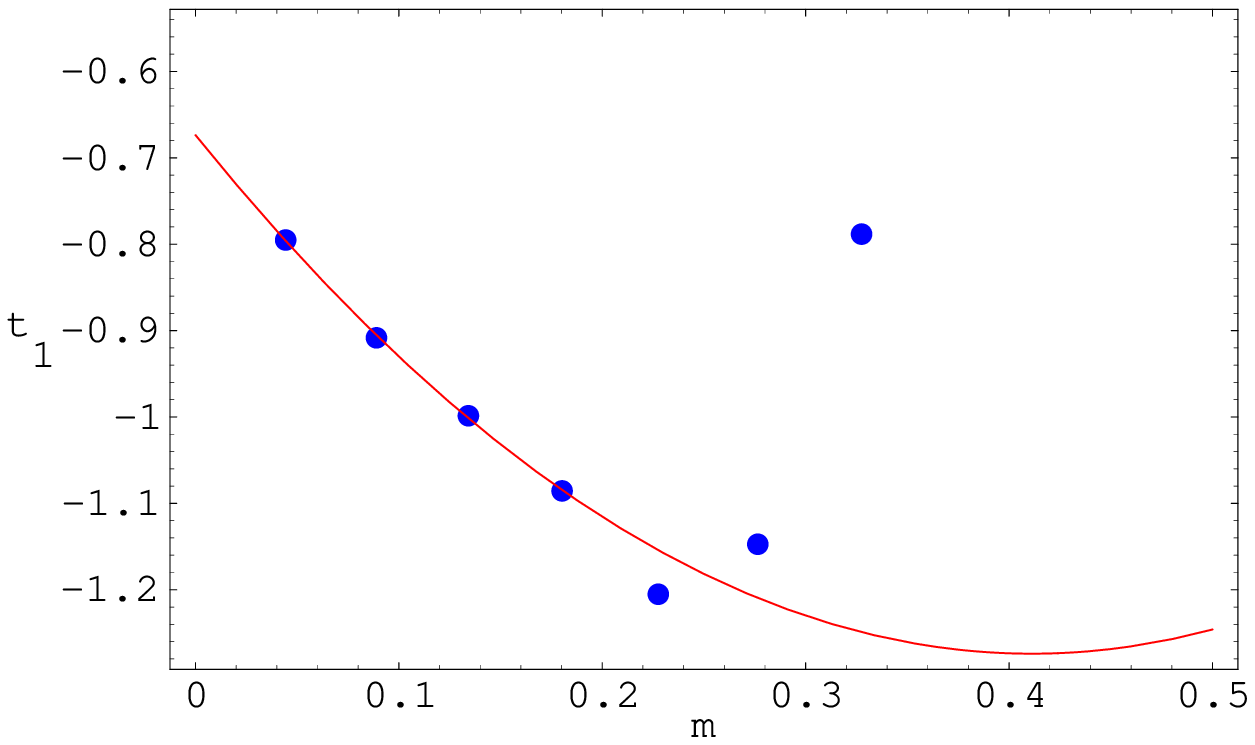 scaled 500}
\BoxedEPSF{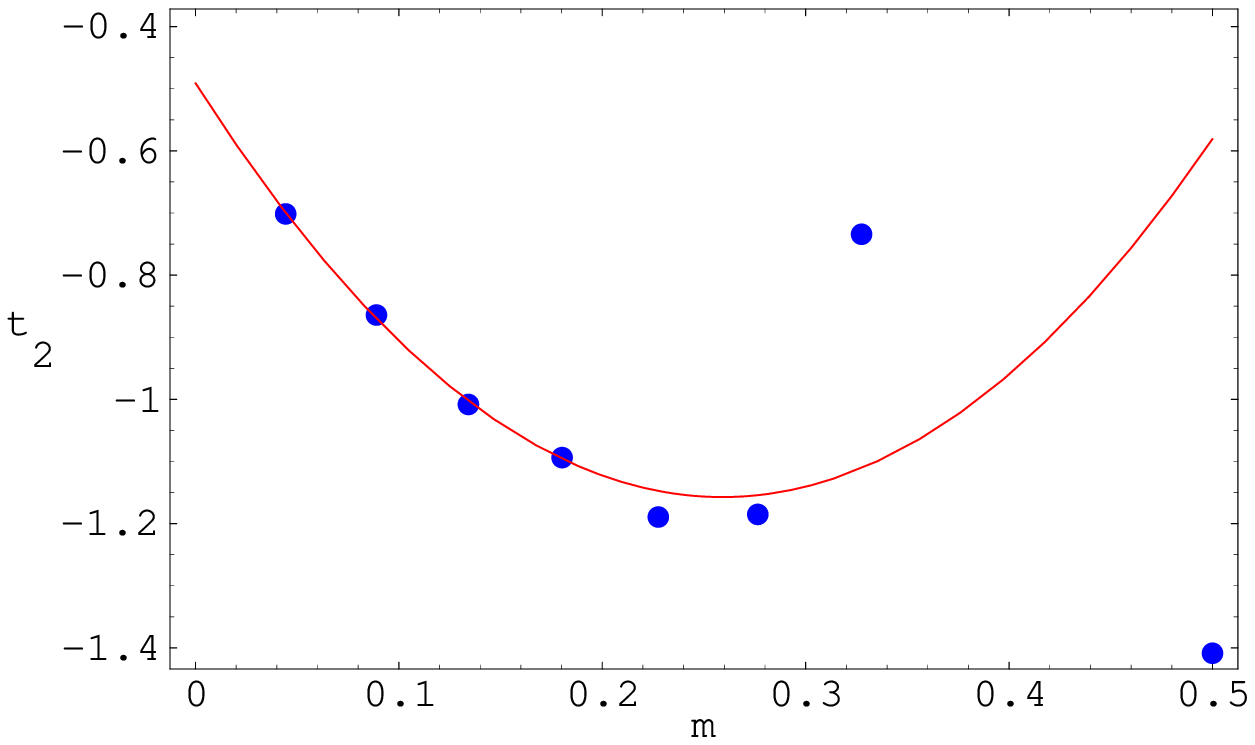 scaled 500}
\caption{The couplings along ${\cal T}_s$. 
\label{traj}}
\end{figure}

\begin{table}
\centering\[
\renewcommand{\arraystretch}{1.25}
\begin{array}{c|cccccccc|c}
 m & l_1 & l_2 & l_3 & l_4 & l_5 & b & t_1 & t_2 & \chi^2 
\\[7.5pt]\hline\hline
0.0444 & 0.012 & -0.068 & 437 & 0.097 & 182 & 0.429 & -0.795
& -0.702 & 13.36\\
0.0890 & -0.005 & -0.075 & 426 & 0.118 & 197 & 0.504 &
-0.908 & -0.864 & 14.28\\
0.1342 & -0.035 & -0.079 & 463 & 0.132 & 211 & 0.575 &
-0.999 & -1.008 & 15.99\\
0.1803 & -0.038 & -0.090 & 374 & 0.158 & 143 & 0.688 &
-1,086 & -1.094 & 18.76\\
0.2275 & -0.037 & -0.101 & 62.4 & 0.212 & 18.8 & 0.858 &
-1.205& -1.189 & 21.14\\
0.2765 & -0.048 & -0.137 & 111  & 0.208 & 22.4 & 0.962 &
-1.147& -1.185 & 25.28\\
0.3275 & -0.007 & -0.144 & 60.7 & 0.163 & 9.6 & 1.010 &
-0.788 & -0.734 & 31.87\\
0.3812 & -0.007 & -0.195 & 12.3 & 0.308 & 2.28  & 1.381 &
-1.207 & -1.302 & 31.83\\
0.4384 & -0.047 & -0.175 & 50  & 0.331 & 12.5 & 1.530 &
0.588 & -0.635 & 43.35\\
0.500 & 0.319 & -0.313 & 822 & 0.367 & 318 & 1.906 &
-1.42& -1.605 & 59.78
\end{array}\]
\caption{The trajectory which minimises the $\chi^2$ test of
Lorentz covariance.
\label{tabtraj1}}
\end{table}

\begin{figure}
\centering
$m$\hspace{0.1in}\BoxedEPSF{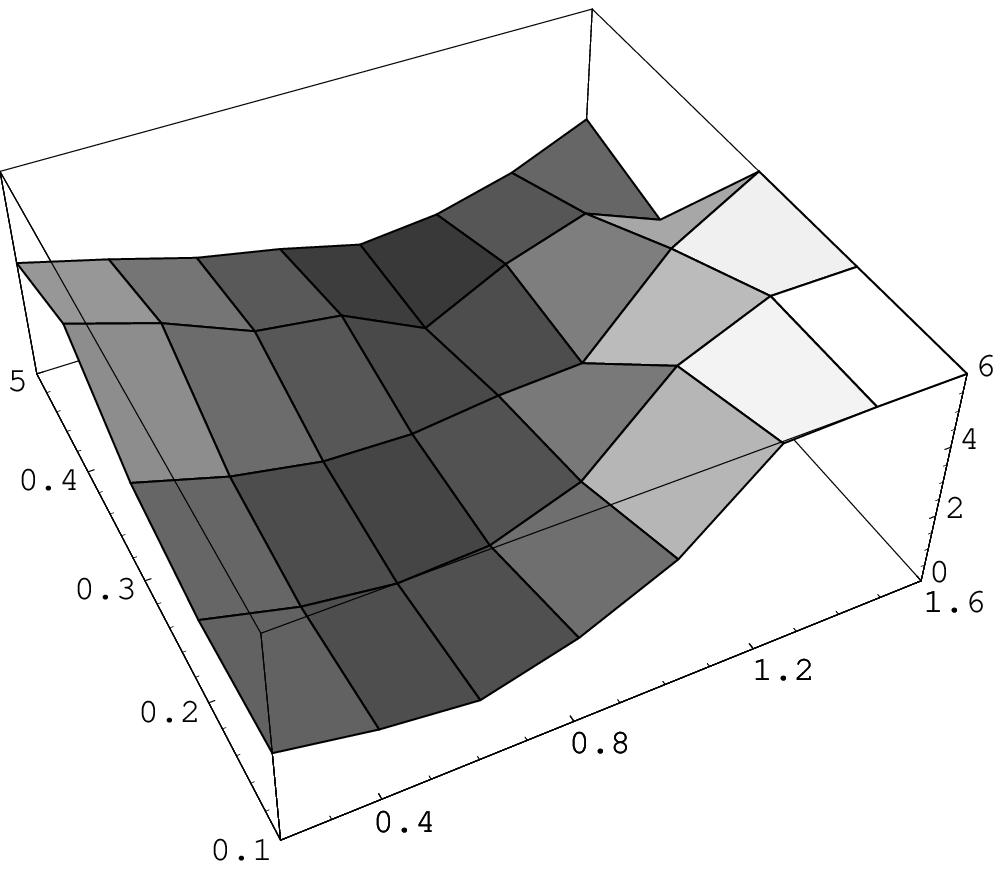 scaled 500}
\BoxedEPSF{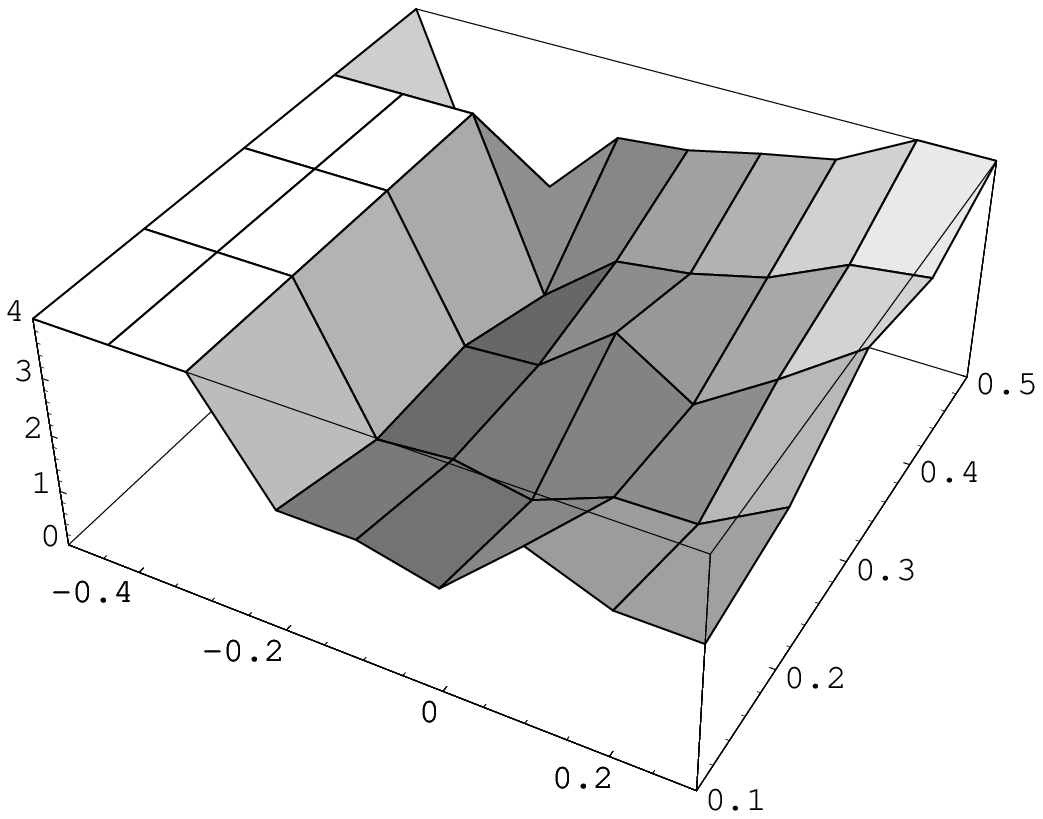 scaled 500}\hspace{0.1in} $m$\\[-20pt]
\hspace{-0.2in} $b$ \hspace{1.9in}  $l_1$\\[12pt]
\BoxedEPSF{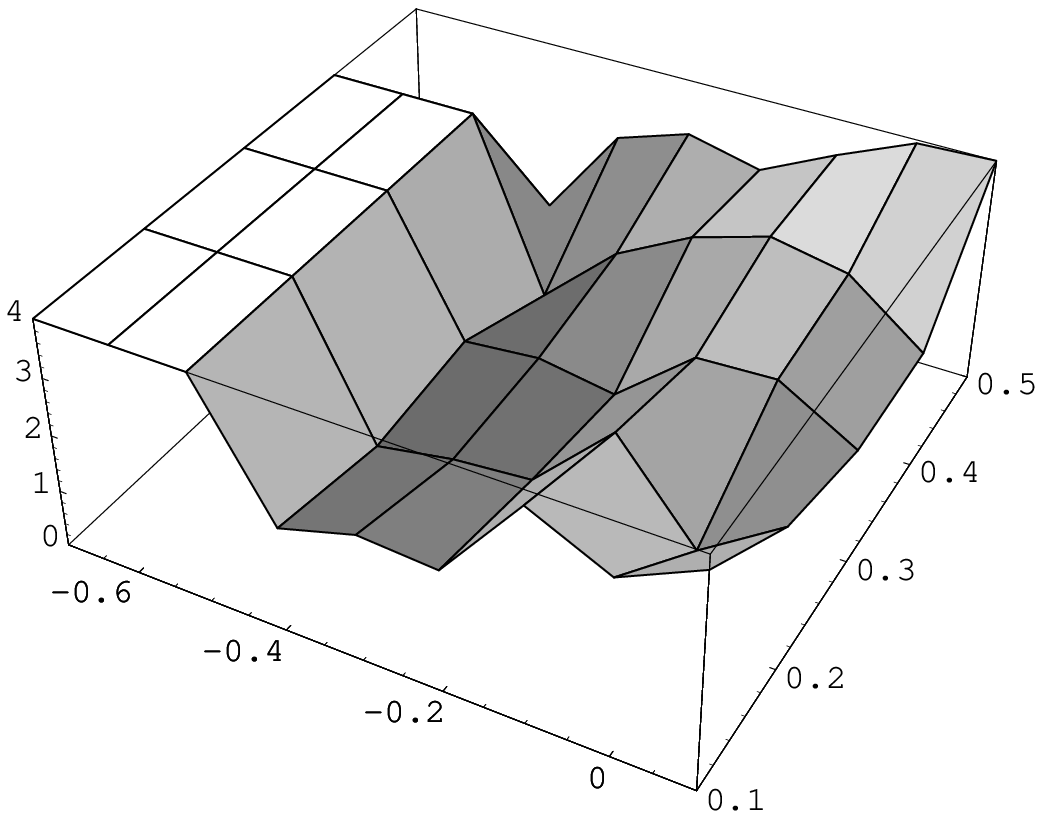 scaled 500}\hspace{5pt}$m$
\hspace{0.1in} \BoxedEPSF{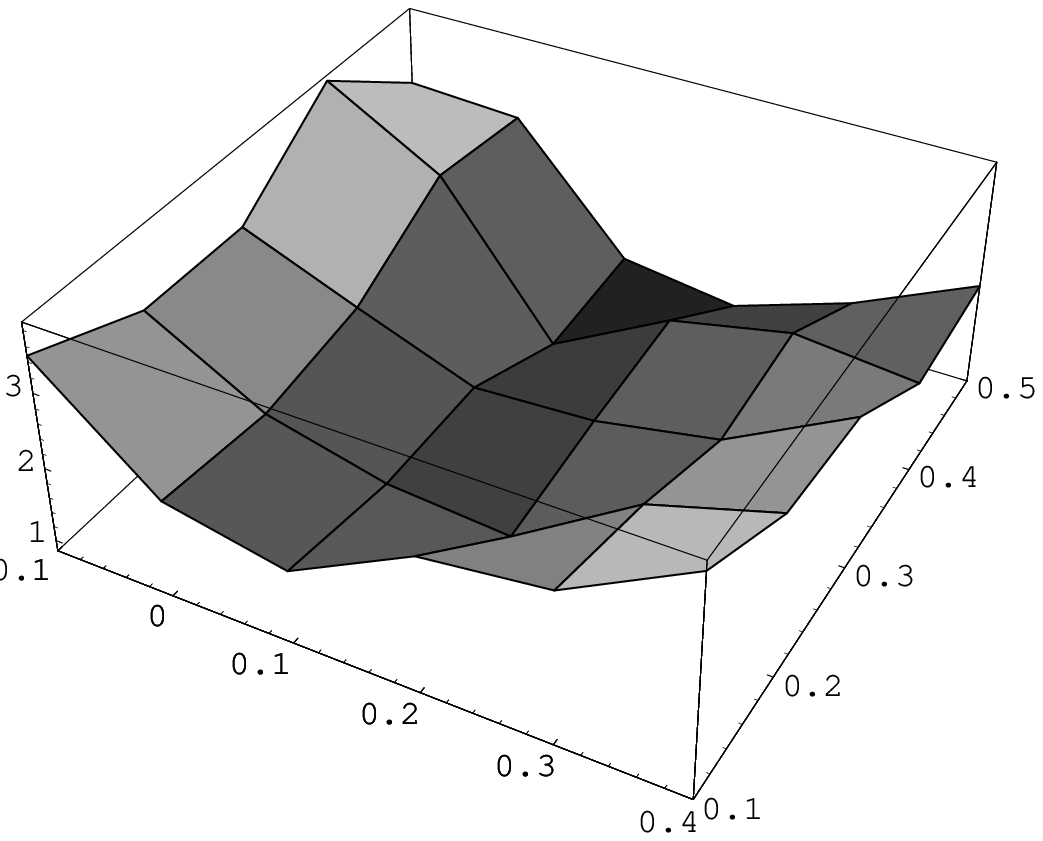 scaled 500} \hspace{0.1in} $m$\\[-20pt]
\hspace{-0.6in} $l_2$ \hspace{2.1in} $l_4$
\\
\caption{Charts of coupling space with $\chi^{2}_{\rm min}$
as the height variable. 
\label{chart}}
\end{figure}

The first step in the process requires some experimentation with
different $\chi^2$ variables.  In general, we used the isotropy of
glueball dispersion relations, rotational invariance of the
heavy-source potential, and Lorentz multiplet structure of the
low-lying glueball spectrum as the ingredients in $\chi^2$.  After
some fine-tuning of weights, we found a very clear signal for a
trajectory ${\cal T}_s$ running from small to large $m$ on which the
$\chi^2$ was significantly reduced to about one per effective degree
of freedom.
Fig.~\ref{traj} and Table~\ref{tabtraj1} show the trajectory 
found by \mi.  Further
details of the $\chi^2$ choice are available in a data file \cite{data}.
40 variables were used in $\chi^2$, though 15 of these were given
significantly more weight than the others, corresponding to
groundstate observables in each symmetry sector. The lowest values of
$\chi^2$, in the range 10--20, occur on a smooth trajectory through the
lowest four non-zero values of $m$ sampled, above which ${\cal T}_s$
begins to break up somewhat.  Fig.~\ref{chart} shows the global
topography of coupling space via \mii, confirming that the
trajectory found in \mi {} sits at the bottom of a well-defined,
unique valley. (Note that regions giving tachyonic glueballs were
assigned $\chi^2 = \infty$ for the purposes of Fig.~\ref{chart}).

\begin{figure}
\centering
$\displaystyle a \sqrt{\sigma}$\BoxedEPSF{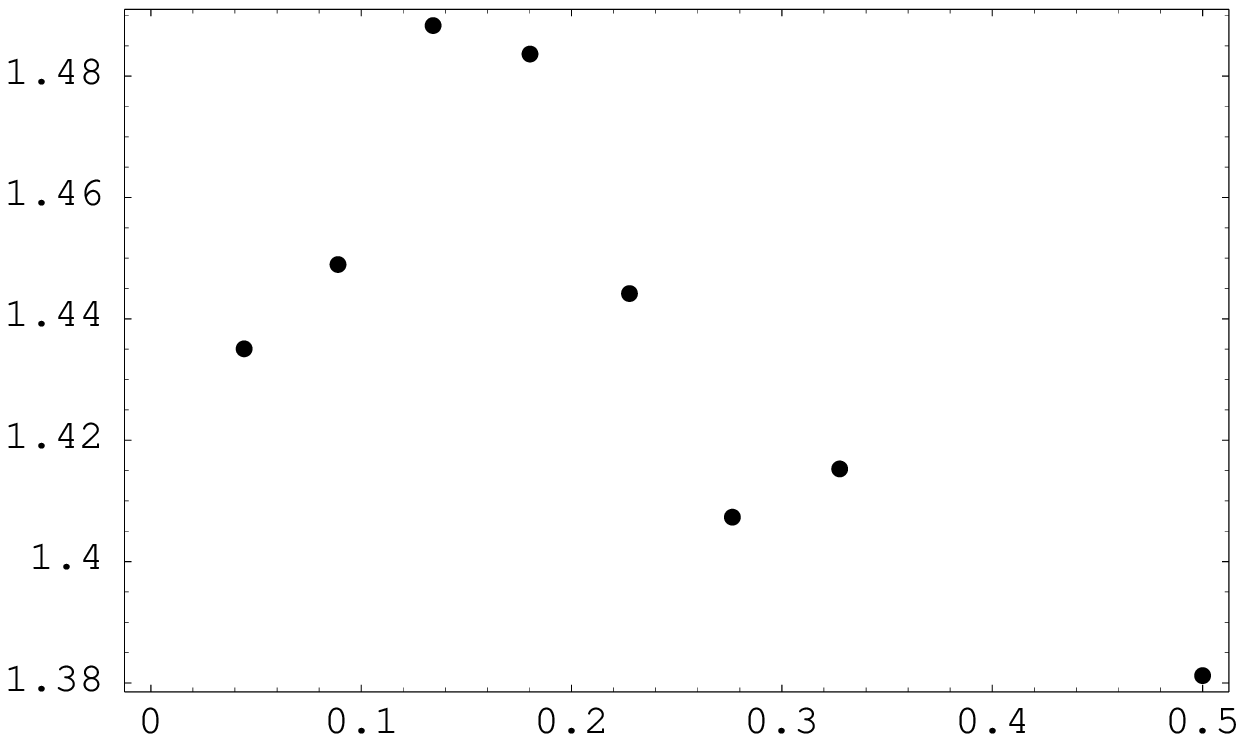 scaled 600}\\
\hspace{0.5in}$m$
\caption{The variation of the lattice spacing $a$ with link-field mass $m$
along ${\cal T}_s$, {\em assuming} exact scaling holds (this only
holds,
even approximately, at small $m$). 
\label{spacing}}
\end{figure}

Physically one expects $m^2 \to -\infty$ as $a \to 0$, since $M$
should be
forced to lie on the $SU(N)$ group manifold $M^{\da} M = 1$ in the
continuum limit. Since we have limited ourselves to the region $m^2 >
0$, we are some way from the continuum. It is remarkable, therefore,
that such a strong signal for the Lorentz trajectory ${\cal T}$ is
obtained.  We can deduce the lattice spacing in units of $\sigma$ on
${\cal T}_s$ if we assume no scaling violations; see
Fig.~\ref{spacing}.  Na\"{\i}vely one expects $m$ to increase with $a$, but
measurements of 
$a$ vs.\ $m$ are very sensitive to scaling violations. 
Fig.~\ref{spacing} is roughly consistent for the smallest
$m$ region, though scaling violations appear much stronger in $3+1$
dimensions than they were in $2+1$-dimensions \cite{dv2}.

\section{Glueball data}
\label{results}
 
\subsection{Masses}

\begin{figure}
\centering
$\displaystyle\frac{\cal M}{\sqrt{\sigma}}$\hspace{0.05in}
\BoxedEPSF{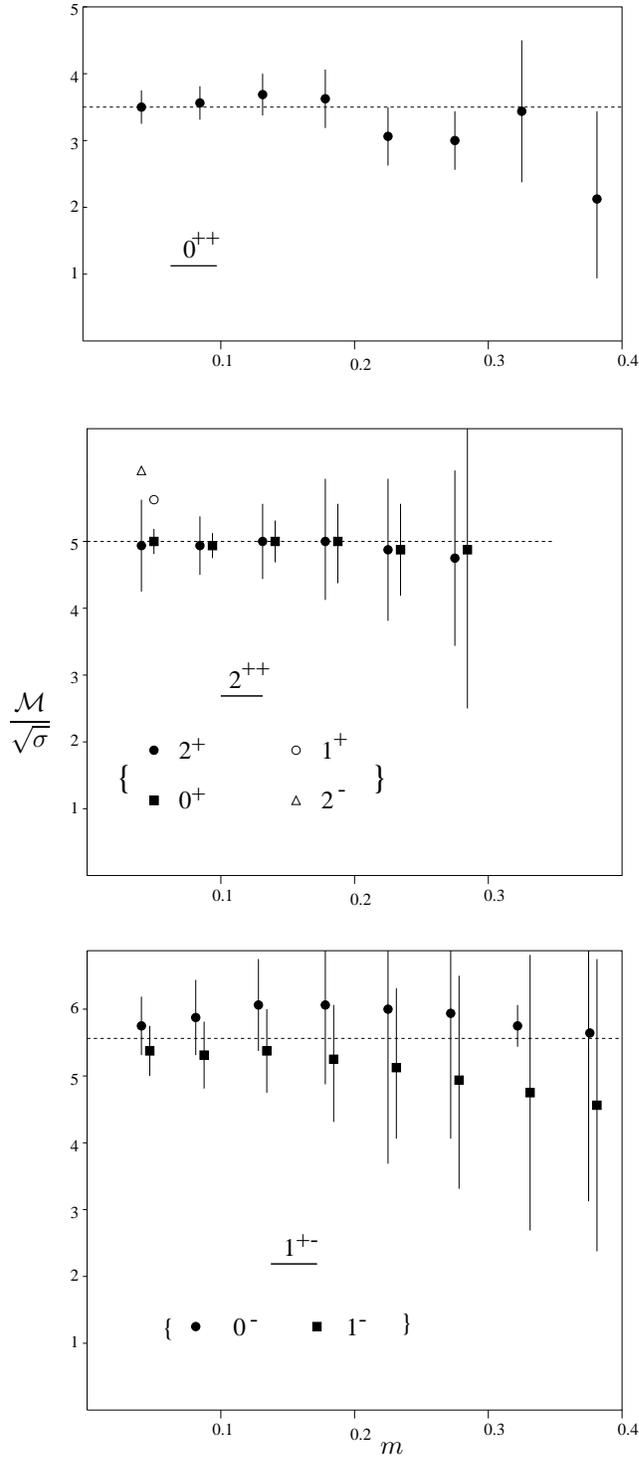 scaled 350}\\
\hspace{0.1in}$m$
\caption{The variation of glueball masses with link-field mass $m$
along
${\cal T}_s$. The open-symbol data for the $2^{++}$, which are
plotted only at $\chi^2 = \chi^2_{o}$, are still too 
inaccurate for error estimates. Data at given $m$ are displaced horizontally
for clarity.
\label{scaling}}
\end{figure}

In the limit ${\cal L} \to \infty$, in addition to 
$x^3$-boost invariance, the transverse lattice has
exact discrete symmetries of the group $D_4$ \cite{bard2}, together with 
charge-conjugation ${\cal C}$. Reflections 
${\cal P}_1$ and ${\cal P}_2$, about $x_1 = 0$ and 
$x_2 =0$ respectively, can be used to  estimate parity ${\cal P}
={\cal P}_1 {\cal P}_2 {\cal P}_3$
from free-field estimates of the light-front parity ${\cal P}_3: x^+
\leftrightarrow x^-$. 
90-degree rotations $x^1 \to x^2$ are exact and
can be used to distinguish the angular momentum
projections ${\cal J}_3 = 0, \pm 1, \pm 2$ from each other. 

In Fig.~\ref{scaling}, we plot the scaling behaviour of the lightest
glueball masses ${\cal M}$ along ${\cal T}_s$,
labelled by $|{\cal J}_3|^{{\cal P}_1}$
and grouped into would-be
Spin--Parity--Charge-Conjugation multiplets ${\cal J}^{\cal PC}$.
We
have estimated error bars on each datum by combining well-understood
(and small) DLCQ and Tamm-Dancoff extrapolation errors ($\sim 0.03
{\cal M}$) with an estimate of systematic finite-$a$ errors coming
from violations of Lorentz covariance, using the formula
\be
{\cal M}(\chi^{2}_{o}) {\chi^2 \over \chi^{2}_{o}}
| c - 1 |_{\rm av.} \ . \label{error}
\eq
$| c - 1 |_{\rm av.}$ is the averaged deviation of the speed of light in 
transverse directions ${\bf x}$, relative to direction $x^3$.
The datum of lowest overall $\chi^2 = \chi^{2}_{o}$ occurs at $m=0.044$.
Formula (\ref{error}), which is supposed to represent roughly
the uncertainty of the intercept
${\cal M}$ of the dispersion relation, works 
well for nearly covariant glueballs in $2+1$-dimensional
studies where the `exact' mass is known \cite{teper1,dv2}. It is also 
consistent with the magnitude of scaling violations and Lorentz
multiplet splittings in the present case.
The various components of a Lorentz multiplet in Fig.~\ref{scaling}
become rapidly more covariant,
measured by their isotropy and degeneracy, as $m$ is reduced.\footnote{ 
Na\"{\i}vely, one might expect the colour-dielectric expansion to break
down at the very smallest $m$, as the link fields become very light. 
However, the energy barrier to production
of more links comes principally from gauge self-energy contributions
at these small masses \protect\cite{bard2, fran1, fran}.}

\begin{table}
\centering\[
\renewcommand{\arraystretch}{1.25}
\begin{array}{|cc|cc|c|}
\hline
\displaystyle {\cal J}^{\cal P C} &|{\cal J}_3|^{{\cal P}_1} &  c_{\rm on} &
c_{\rm off}  & \displaystyle{{\cal M} \over \sqrt{\sigma}} \\ \hline \hline
0^{++} & 0^{+}  & 1.07 & 1.07 & 3.50(24) \\ \hline
2^{++} & 2^{+}  & 0.86 & 0.87 &  4.97(68)   \\ 
       & 2^{-}  & {\rm Im} & - & 6.07(?)  \\ 
       & 1^{\pm}& 0.91,{\rm Im} & 0.72, {\rm Im} & 5.62(?) \\ 
       & 0^{+}  & 0.99 & 0.98 & 4.97(18)   \\
\hline 
1^{+-} & 1^{\pm} &  0.88, 1.06 & 
                    0.94, 1.00  & 5.37(36)  \\ 
       & 0^{-} & 1.07 & 1.07 & 5.74(45) \\
\hline     
\end{array}\]
\caption{The glueball multiplet components at the
lowest overall $\chi^2 = \chi^2_{o}$, showing 
masses from ${\cal M}^2 = 2P^+ P^- ({\bf P} = 0)$ and 
$c$ from  the dispersion relation (\protect\ref{latshell}). 
``Im'' indicates $c^2 < 0$
and ``--'' indicates
the quantity has not been measured.
The ${\cal J}_3 = \pm 1$ states are exactly degenerate. 
\label{table1}}
\end{table}

The final mass estimates given in the introduction are displayed in
more detail in Table~\ref{table1}.  They are taken from the point of lowest
$\chi^2 = \chi^{2}_{o}=13.36$ on the scaling trajectory.  
We are confident that the value of and
error on these results is a reasonable estimate of the continuum
result, provided the relevant observable exhibits a scaling region
along ${\cal T}_s$ and the error (\ref{error}) is small. We emphasise
that scaling alone is not sufficient for a good estimate of the
continuum result.  Table~\ref{table1} gives some idea of where the
remaining errors lie. In particular, the ${\cal P}_1 = -1$ components
of the tensor glueball are still dominated by lattice artifacts,
requiring further terms in the Hamiltonian presumably. Our lightest
$0^{-+}$ state lies at $\sim 8 \sqrt{\sigma}$, compared to much lower
$SU(3)$ Euclidean lattice Monte Carlo results \cite{star}; but we have
no measure of its covariance and suspect a very large error on our
result.

It is interesting to plot glueball masses in pure gauge theory versus
$1/N^2$, since this is supposed to be the relevant expansion parameter
about $N= \infty$ \cite{hoof}; see Fig.~\ref{alln} in the
introduction.  There is remarkably little variation of glueball masses
with the number of colours.  It gives further support to the notion
that $N=3$ is close to $N=\infty$ in many situations.  In particular,
popular flux-tube \cite{paton} and string models \cite{nambu} of the
soft gluonic structure of hadrons are typically more appropriate to
the large-$N$ limit of QCD (or are independent of
$N$). Figure~\ref{alln} indicates that these models should give
worthwhile quantitative 
approximations to $N=3$ pure QCD .  One can even go so far
as to derive a formula for glueball masses at any $N$, by fitting
those at $N=2,3, \ldots, \infty$ to polynomials in $1/N^2$; this might
be a useful guide for higher-$N$ ELMC studies \cite{teper2}.

\subsection{Covariance}

\begin{figure}
\centering
$\displaystyle\frac{V_{QQ}}{\sqrt{G^2 N}}$\BoxedEPSF{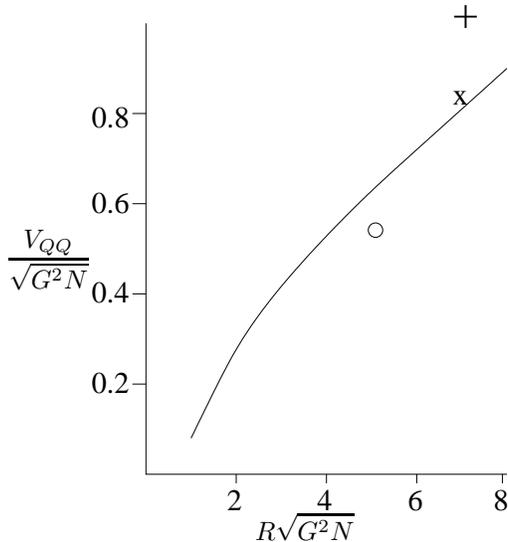 scaled 600}\\
\hspace{0.5in}$\displaystyle R \sqrt{G^2 N}$
\caption{Heavy-source potential $V_{QQ}$ vs.\ source separation
 $R = \sqrt{(\delta x^3)^2
+ (\delta x^2)^2 + (\delta x^1)^2}$. Shown is a fit to the measured
potential at zero transverse 
separations $\delta x^1 = \delta x^2 = 0$. 
The `off-axis' data are: (O)  $\delta x^3 =1.22$, $\delta x^2 = 4.95$,
$\delta x^1 = 0$; (X)  $\delta x^3 =4.89$, $\delta x^2 = 4.95$,
$\delta x^1 = 0$; (+)  $\delta x^3 =0$, $\delta x^2 = 4.95$,
$\delta x^1 = 4.95$. Units are $(G^2 N)^{-1/2}$. 
\label{rond}}
\end{figure}

In searching for ${\cal T}_s$ we used data from the heavy-source
potential and glueball dispersion near ${\bf P}=0$ (the speed of
light). At the $\chi^2= \chi^{2}_{0}$ point on ${\cal T}_s$, the
potential is plotted in Fig.~\ref{rond}.  The potential between
sources separated by distance $R$ only in the $x^3$ direction has been
fit to 
\be 
{V_{QQ}\over \sqrt{G^2 N}} = 0.084 R\sqrt{G^2 N} + 0.259 -
{0.265 \over R \sqrt{G^2 N}} \label{poten} \ .  
\eq 
The remaining
points in Fig.~\ref{rond} are `off-$x^3$-axis' samples of the
potential at a general separation $R$, whose deviations from the value
given by (\ref{poten}) are used in the $\chi^2$ test. As can been
seen, the rotational invariance of the potential is patchy, showing
significant deviations at very small $x^3$ separations and/or along the
$(1,1)$ transverse direction (see discussion in Section~\ref{heavy}).

\begin{figure}
\centering
$\displaystyle\sqrt{\frac{2P^+ P^- - {\cal M}^2} {
\sigma}}$\BoxedEPSF{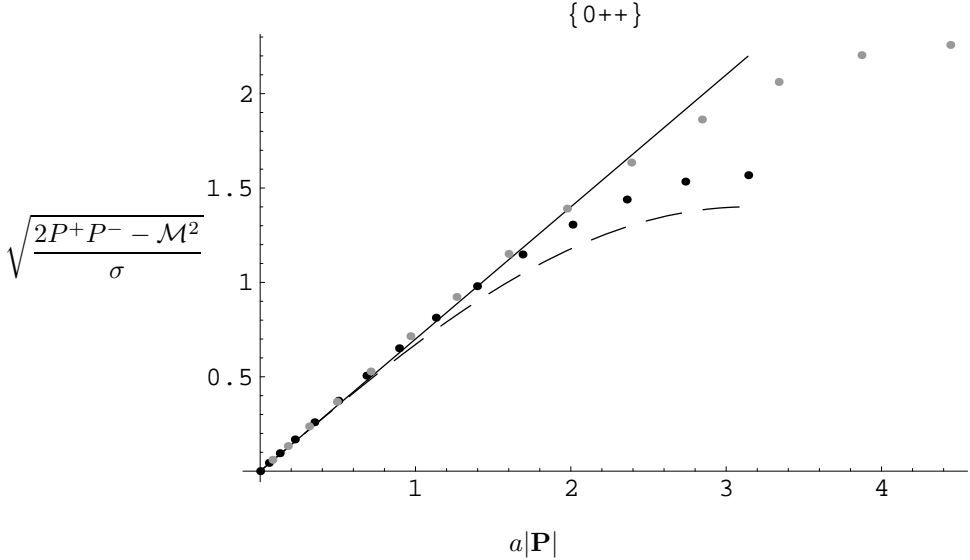 scaled 1000}\\[-40pt]
\hspace{0.5in}$a | {\mathbf P} |$
\caption{Dispersion relation of $0^{++}$ glueball. Straight line is
exact relativistic form, dots are data at $\chi^2= \chi^{2}_{0}$ point
on ${\cal T}_s$ (black for direction ${\bf P} \propto (1,0)$,
gray for ${\bf P} \propto (1,1)$). The dashed line is a `lattice'
dispersion $\propto \sqrt{1-\cos{a |{\bf P}|}}$ along
 ${\bf P} \propto (1,0)$, normalised
to the correct slope at ${\bf P}=0$.
\label{curve}}
\end{figure}

Only the long range form of the glueball dispersion was used as a
variable in the $\chi^2$ test.  As a further consistency check, we can
also plot the dispersion throughout the Brillouin zone $0 < | {\bf P}
| < \pi/a$.  As the example of the lightest $0^{++}$ glueball
illustrates (Fig.~\ref{curve}), improvement towards the full
relativistic form seems to occur on ${\cal T}_s$.

\subsection{Wavefunctions}

\begin{table}
\centering\[
\renewcommand{\arraystretch}{1.25}
\begin{array}{c|ccccccc}
& {\rm scalar} & & & \hspace{-0.4in} {\rm tensor}  & & & 
\hspace{-0.4in}{\rm vector} \\
{\rm Shape}  & \{ 0^+ \} & \{ 2^+  & 2^- & 1^{\pm} & 0^+ \} & \{ 1^{\pm}  &
0^- \}  
\\[7.5pt]\hline\hline
m   & 0.019 & 0.016  & 0  & 0  & 0.014 &  0.810   & 0     \\
l_2 & 0.003 & 0.518  & 0  & 0  & 0.535 &  0.039  & 0    \\
l_1 & 0.002 & 0.369  & 0  & 0 & 0.361 &  0.016   & 0    \\
l_4 & 0.483 & 0  & 0.882  & 0.923  & 0.017 &  0.052   & 0.049   \\
b   & 0.400 & 0  & 0.024  & 0  & 0.003 & 0.069    & 0.860    \\
{\rm other}  & 0.093  & 0.097  &  0.094  & 0.077   & 0.070 & 0.014  &
0.091  \\
\end{array}\]
\caption{Probability distribution of transverse 
shapes  for various $|{\cal J}_3|^{{\cal P}_1}$ components of
the low-lying glueball eigenfunctions (following Table~\ref{table1}). 
The shapes are denoted by the
coupling constant for that shape in $P^-$; 
for example, $m$ denotes the
$\Tr \left\{ M_r M_{r}^{\da} \right\}$ operator. Contributions $< 0.001$ are
neglected.
\label{shape}}
\end{table}

From the explicit glueball eigenfunctions one can extract various
measurements of glueball structure.\footnote{The results in this
subsection are from an earlier calculation that used a slightly
different choice of $\chi^2$ test than that appropriate to
Table~\ref{tabtraj1}. The difference in the resulting wavefunctions is
negligible for the purposes of our discussion in this subsection.}
While these results are still somewhat academic until one couples
quarks to allow the glueballs to decay, they are still important. One
can check whether the constituent structure of boundstates
is valid. The probability distribution of transverse shapes is a useful
measure of this (see Table~\ref{shape}). If eigenfunctions were to
contain large mixtures of different numbers of link-partons, it would
call into question the validity of the colour-dieletric expansion.

\begin{figure}
\centering
\begin{tabular}{c@{}c}
      & $0^{++}$\\[5pt]
$G_d$ & \BoxedEPSF{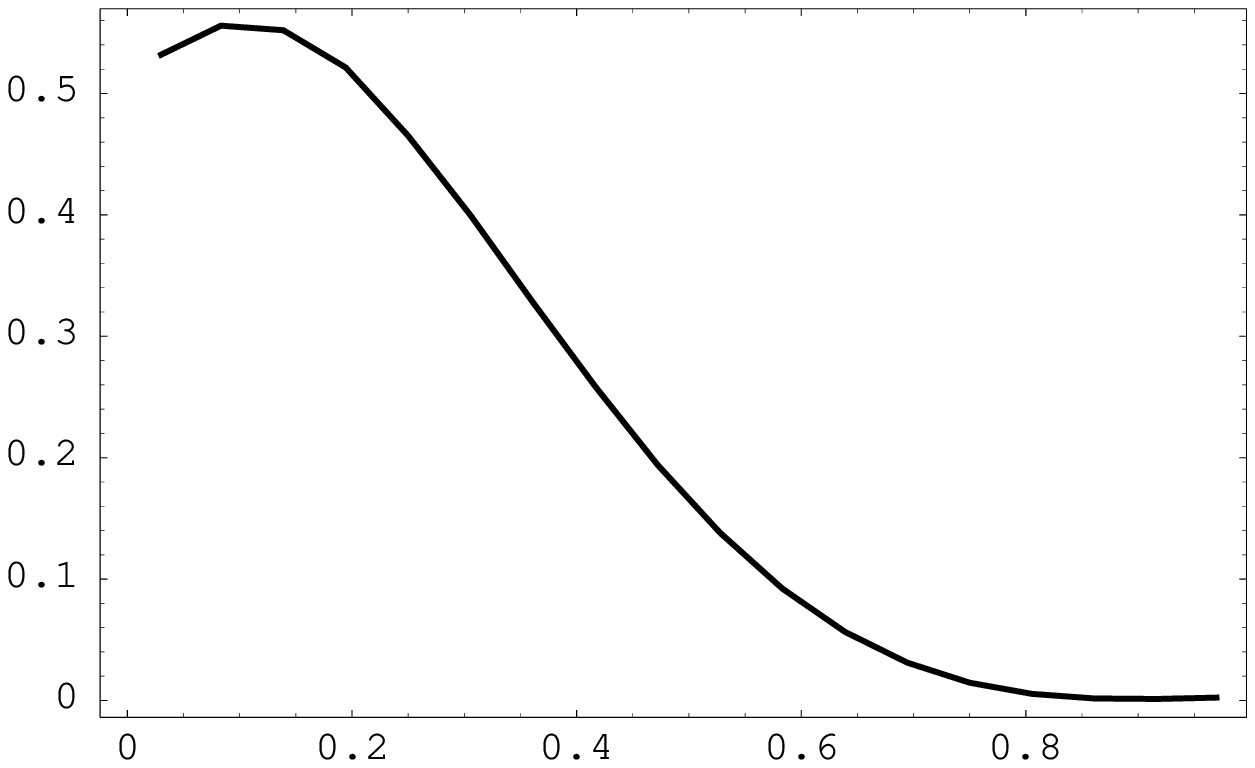 scaled 600}\\
      & $x$ \\[20pt]
      & $2^{++}$ \\[5pt]
$G_d$ & \BoxedEPSF{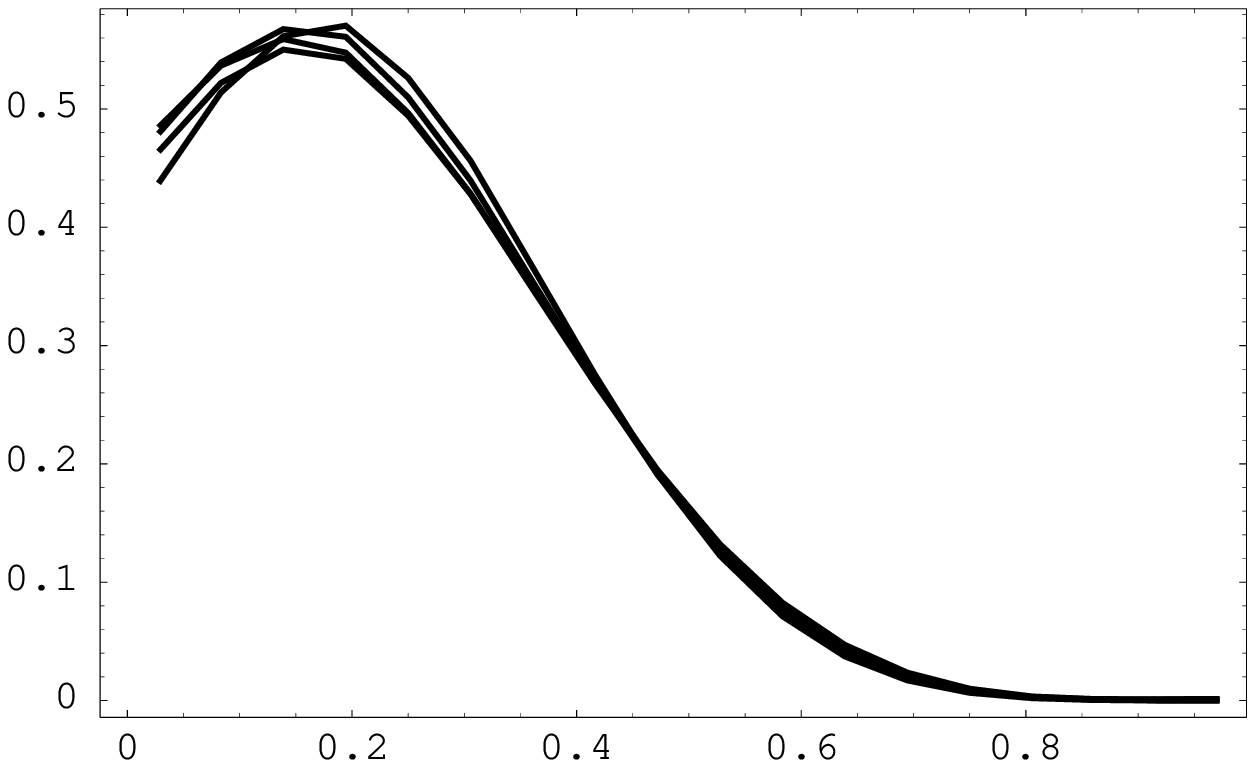 scaled 600}\\
      & $x$\\[20pt] 
      & $1^{+-}$ \\[5pt]
$G_d$ & \BoxedEPSF{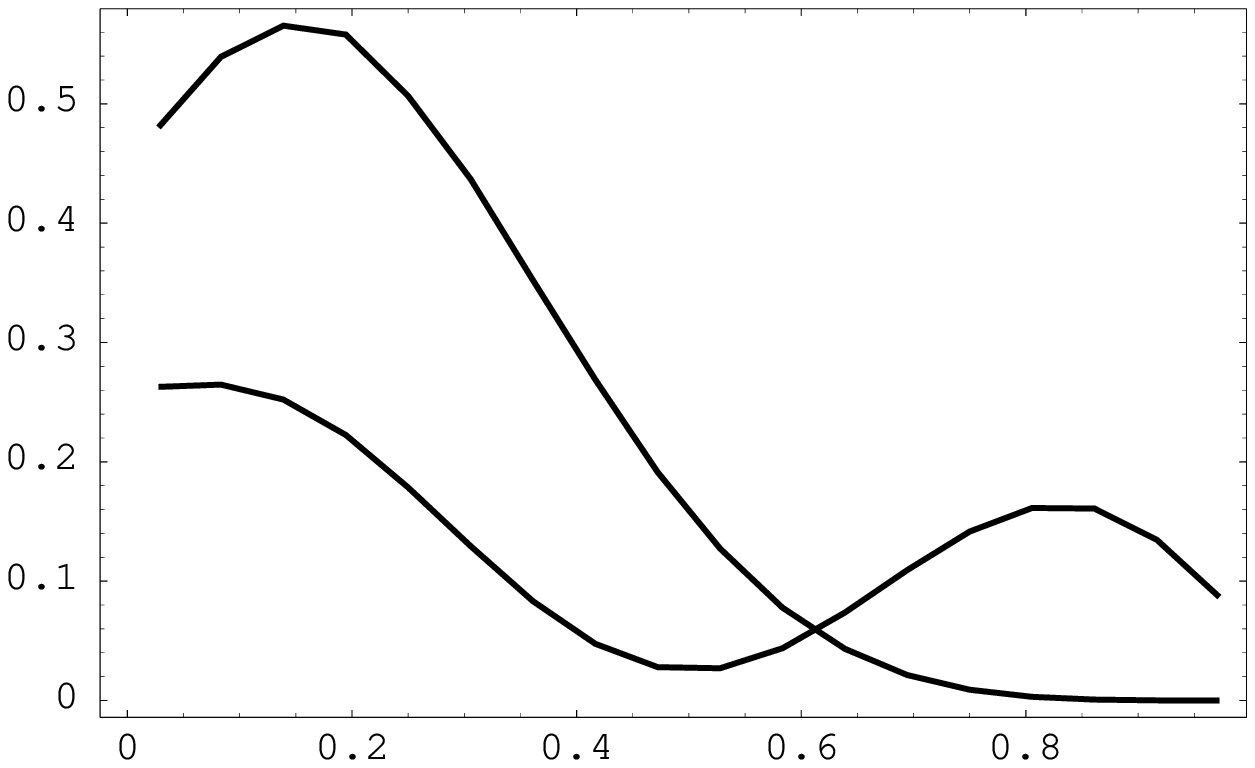 scaled 600}\\
      & $x$
\end{tabular}
\caption{The distribution of $P^+$ momentum in various ${\cal J}_3$
components of the low-lying glueballs. 
\label{dist}}
\end{figure}

Another interesting quantity is the distribution of longitudinal
momentum $P^+$ among the link partons. In Fig.~\ref{dist} we plot the
quantity
\begin{eqnarray}
G_{d}(x,1/a)  & = & 
{1 \over 2\pi x P^+} \int dx^- {\rm e}^{-{\rm i} x P^+ x^-}
\langle\Psi(P^+)|
        \Tr\left\{ \partial_{-} M_r \partial_{-} M^{\dagger}_r \right\}
|\Psi(P^+)\rangle \ ,
\end{eqnarray}
which measures (in $A_- = 0$ gauge)
the probability of finding a link-parton carrying
momentum fraction $x$ of the glueball momentum $P^+$.  It depends upon
the transverse normalisation scale through $a$, though our piece of
the Lorentz trajectory is too short to reliably see physical evolution
of $G_{d}$ with scale. $a \sim 0.65\, {\rm fm}$ for the data shown.

$G_{d}$ is related to the gluon distribution; it becomes the gluon
distribution in the limit $a \to 0$. Moreover, since $M$ is some
collective gluon excitation and the momentum sum rule is satisfied,
one would na\"{\i}vely expect the gluon distribution at a general scale $a$
to be softer than $G_{d}$. In this case, the $0^{++}$ glueball
(Fig.~\ref{dist}) does not resemble gluonium (two-gluon boundstate),
which would have a distribution symmetric about $x=0.5$.  The fact
that all components of the tensor's distribution look like the
$0^{++}$ is also interesting. It may indicate that the tensor's
angular momentum comes mostly from gluon spin; however we caution that
not all components of our tensor are yet behaving in a covariant
fashion.  Once light quarks are coupled to the problem, these
distributions should have distinctive experimental signatures.

\section{Conclusions}
\label{conclusions}

In this paper we have carried through the program, first
suggested by Bardeen {\em et al.}\ \cite{bard2}, for solving the
boundstate problem of pure QCD on a transverse lattice. Three
of our developments
were crucial to this success:
inclusion of higher Fock sectors via DLCQ; 
inclusion of the full set of allowed operators occurring to lowest order
of the colour-dielectric expansion of the light-front
Hamiltonian; efficient semi-analytic methods for computing and
extrapolating  many-body
light-front Hamiltonians \cite{dv1,dv2}. In addition, we combined the analysis 
of the pure glue sector with an analysis of the heavy
source sector, which is needed not only to provide an
absolute scale but also to pin down the correct Lorentz-covariant
Hamiltonian in each sector. The upshot of all these
developments is that we were able to clearly identify a scaling trajectory 
in the space of couplings on which Lorentz covariance of observables is
enhanced. 

We believe that this trajectory is an approximation to 
the exact Lorentz-covariant, scaling trajectory in the
infinite dimensional space of all Hamiltonians.
Strong evidence in favour of this comes from the
scaling behaviour of glueball masses, whose values agree with
continuum results (extrapolated to large-$N$) from established
methods. It is largely straightforward to relax the approximations
we made so far, including the large-$N$ limit,
to increase the scope and accuracy of our results.

The light-front wavefunctions are entirely new results.
Their accuracy is largely undetermined and, in their present form,
they cannot easily be used to compute an experimentally measured
quantity. However, their interesting structure might provide a source
for phenomenological models. In fact, given the accuracy of
the glueball masses and
covariance of the dispersion relations, we would be very surprised to
find that they were inaccurate.

To make contact between light-front wavefunctions and the wealth
of hadronic experimental data,
the next step must introduce finite-mass propagating quarks. 
Although overviews of the standard Kogut-Susskind \cite{bard1} and
Wilson \cite{matthias} fermions on a transverse lattice have been given, no
first-principles calculations in the manner of this paper have
been completed. By adding observables that test chiral symmetry 
(and parity) we can hope to find a similar scaling trajectory. 
Such calculations are now underway.

\vspace{10mm} 
\noindent Acknowledgements: We thank
our colleagues in DAMTP, CERN, Erlangen and the ILCAC and especially
M. Teper for help and encouragement at various stages.
SD is supported by PPARC grant No.\ GR/LO3965.

\end{document}